\documentclass{statsoc}

\usepackage[T1]{fontenc}
\usepackage[utf8]{inputenc}
\usepackage{amsmath,amssymb}
\usepackage{mathrsfs}
\usepackage{float}
\restylefloat{table}

\usepackage{geometry}               
\usepackage{enumerate}  
\usepackage{setspace}
\usepackage{graphicx}
\usepackage{tabularx}
\usepackage[parfill]{parskip}
\usepackage{epstopdf}
\usepackage{subfigure}
\usepackage[ ruled ]{algorithm2e}
\usepackage[toc,page]{appendix}
\usepackage{changepage}  

\usepackage{color,soul}

\newcommand\scalemath[2]{\scalebox{#1}{\mbox{\ensuremath{\displaystyle #2}}}}

\title[Deconvolution of dust mixtures by latent Dirichlet allocation]{Deconvolution of dust mixtures by latent Dirichlet allocation in forensic science}
\author[M. A. Ausdemore {\it et al.}]{M. A. Ausdemore and C.Neumann}
\address{Department of Mathematics and Statistics, South Dakota State University, Brookings, SD, USA}
\email{madelineausdemore@gmail.com}

\begin{document}

\begin{abstract}
Dust particles recovered from the soles of shoes may be indicative of the sites recently visited by an individual, and, in particular, of the presence of an individual at a particular site of interest, e.g., the scene of a crime. By describing the dust profile of a given site by a multinomial distribution over a fixed number of dust particle types, we can define the probability distribution of the mixture of dust recovered from the sole of a shoe via Latent Dirichlet Allocation. We use Variational Bayesian Inference to study the parameters of the model, and use their resulting posterior distributions to make inference on (a) the contributions of sites of interest to a dust mixture, and (b) the particle profiles associated with these sites. 
\end{abstract}

\pagebreak

\section{Introduction}
\label{intro}

Dust particles recovered from beneath the soles of an individual's shoes consist of a mixture of dust particles collected from different sources and may be indicative of the locations recently visited by that individual. In particular, this dust may reveal his presence at a location of interest, e.g., the scene of a crime. The contributions of these locations to the mixture may vary as a function of the amount of dust present at each location, the time spent by the individual at each location, the activity of the individual at each location, or how recently the individual visited each location. The profile of a given source of dust can be described by a multinomial distribution over a fixed number of dust particle types, which enables us to describe the mixture of dust recovered from the sole of a shoe by latent Dirichlet allocation \citep{Blei:2003}.
 
In this paper, we describe an algorithm that resolves mixtures of dust to address two different questions of forensic interest. Given a set of samples recovered from one or more objects of forensic interest that consists in mixtures of dust from $M=Q+K$ sources, and samples of dust from $K$ known sources, we are interested in:
\begin{enumerate}[(a)]
	\item inferring the dust profiles of the $Q$ unknown sources;
	\item inferring the proportions in which dust from each of the $Q+K$ sources are present in the samples.
\end{enumerate}
An example of the first inference question may arise when, given an individual suspected of kidnapping with known home and workplace, we are interested in providing information on the dust composition of the unknown location where the victim is being held. Provided that the necessary data and a suitable inference framework exist, the second inference question may be useful to discuss issues such as (a) how long a person stayed at each location, (b) how recently the person visited each location, or (c) what type of activity the person had at each location.

We use latent Dirichlet allocation (LDA) \citep{Blei:2003} to define the generative process that produces mixtures of dust samples. We use variational Bayesian inference (VBI) \citep{Hoffman:2013, Blei:2016} to study the parameters of our model and to address these two inference questions. 


Currently, the use of dust evidence is anecdotical and is limited to cases where rare and characteristic particles are observed (e.g., pollen, seeds, spores - see \cite{Bull:2006, Mildenhall:2006, Mildenhall:2006b, Stoney:2011, Bryant:2014, Stoney:2014} for a discussion of these methods). While most evidence types considered by forensic scientists result from the interactions between criminals and objects or victims at crime scenes, dust evidence arises from the mere presence of individuals at locations of interest, and therefore does not depend on the activity or actions that occur between criminals and objects or victims at the location of interest to be observed. Thus, the goal of this paper is to explore the statistical foundations of a new paradigm for the contribution of forensic science to criminal investigations.


\section{Applying LDA to mixtures of dust}
\label{LDA_Dust}

Topic models aim at discovering hidden semantic structures in a body of documents by grouping together words that are likely to have originated from the same themes or authors. By generalising this concept, LDA can be extended to arbitrary mixtures of objects represented by categorical random vectors, such as particle types. 

Dust sources generate particles that can be observed at different locations, such as a crime scene or the house of a suspect, or one or more objects of forensic interest, such as shoe soles. The model presented in this paper uses dust samples collected from relevant geographical locations and from the surfaces of objects of forensic interest to make inferences on the dust profiles at the different locations, and on their respective contributions to the dust mixture observed on the forensic objects. 

To help describe our research questions, the following parallel between topic modelling and the dust problem can be made:

\begin{adjustwidth}{1cm}{1cm}
	Several authors, each one specialised in a single topic, jointly contribute to a book. We do know the speciality of each author, but we do not know what each topic looks like in terms of the proportions of the different words in the dictionary that are used in each one of them. We are interested in inferring the respective contribution of each author to the book. If we can obtain single-topic documents from all authors, the model can learn the topics from the single-topic documents, and then resolve the mixture of topics in the book. If we can only obtain single-topic documents from a subset of the authors, the model can learn some of the topics from the single-topic documents, and then learn the remaining topics, as well as the contribution of all authors, from the book.
\end{adjustwidth} 

This scenario shows that our inference questions are somewhat different than more traditional topic-models. We are less concerned by what the different topics look like, as opposed to their mixture proportions in a given document. In addition, we have the possibility to infer some of the topic profiles by observing single-topic documents. That said, the number of authors (and therefore the number of topics) is not always known, and by extension, what their corresponding topics look like. When framed in the context of the forensic analysis of dust, the above scenario can be rewritten as:

\begin{adjustwidth}{1cm}{1cm}
	Several geographical locations, each one receiving dust from a single source, were visited by a shoe. We do know that each location is only associated with a single dust source, but we do not know the dust profiles of these locations. We are interested in inferring the respective contribution of each location to the trace dust mixture observed on the surface of the sole of the shoe. If we can obtain dust samples from each location, the model can learn the dust profiles of the locations from these samples, and then resolve the mixture of dust under the shoe. If we can only obtain dust samples from some of the locations, the model can learn some of the dust profiles from the dust samples directly obtained at the locations, and learn the remaining profiles, as well as the respective contribution of the different locations, from the dust mixture recovered on the shoe.\footnote{Appendix~\ref{terms} provides a glossary of terms that connects the vocabulary presented for dust modelling to the more familiar vocabulary of topic modelling to further assist the reader with our development.}
\end{adjustwidth}

In the dust scenario, we make a distinction between the \textit{location} where dust can be sampled, and the \textit{source} from which the dust originates. Our terminology further subsets locations by differentiating \textit{geographical locations}, which correspond to any place that an individual might have visited and where dust might be sampled from, from \textit{trace locations}, which correspond to a location or object where evidentiary dust samples are collected. Critically, our model considers that all the dust at a given geographical location originates exclusively from a single source, while the dust observed on the surfaces of trace objects originates potentially from more than one source (see assumptions (d) and (e) below). This constraint allows for learning the dust profiles of the different geographical locations that might have been visited by an individual and provides a basis to determine if dust from any of these locations is present in the mixture observed on the trace object. 

Our model is different from the original LDA model proposed by \citet{Blei:2003} in that these authors consider that a corpus consists in multiple documents, which are all composed of a mixture of the same topics in the same proportions. In our application, we consider that a corpus consists in multiple documents composed of the same topics, but in varying proportions. In other words, our corpus consists in multiple dust samples: some originating from geographical locations that are known to have been visited by the evidentiary objects (thus, representing single-source dust profiles), and some originating from trace locations (thus, representing mixtures of dust profiles, whose mixture proportions may be different from one evidentiary object to another). This implies that we consider that the parameters that control the contributions of the different dust sources to the dust samples and the parameters that control the particle profiles of these sources are distributed according to asymmetric Dirichlet distributions. In our implementation, these hyper-parameters are represented by matrices rather than vectors.


Our application also differs from supervised topic models \citep{Blei:2007S, Lacoste-Julien:2008, Wang:2009, Zhu:2012}, given that we are interested in determining the relative contributions of the $Q+K$ sources to the mixtures of dust observed in a sample, rather than associating a new set of dust samples with a single specific source. In fact, our problem cannot simply be framed as a supervised learning problem: the number of locations visited by a shoe, or the dust profiles of geographical locations that cannot be studied directly, are unknown and must be inferred from the data.

Finally, our problem differs from that addressed by author-topic models \citep{Rosen-Zvi:2004, Steyvers:2004, Rosen-Zvi:2010}. Although we have made an analogy between authors and geographical sampling locations, our inference questions diverge. The author-topic models proposed by \cite{Rosen-Zvi:2004, Rosen-Zvi:2010} and \citet{Steyvers:2004} aim at discussing which topics are preferred by each author in a fixed set of known authors by assuming a uniform contribution of each author to a single document. The main inference question of our application is the exact opposite: we are interested in studying the respective contribution of each author to a set of documents, given their preferred topics.

\subsection{Model assumptions}
\label{model_assumptions}
To develop our model and account for the differences discussed above, we make the following assumptions:
 \begin{enumerate}[(a)]
 	\item Observations made on particles within a dust sample are exchangeable (\cite{Robert:2007}, page 159). 
 	\item Observations made on dust samples collected at a given location are exchangeable.
 	\item Sources yield dust with a fixed and constant profile. Dust sources do not cross-contaminate each other. 
 	\item The composition of dust samples recovered at a geographical location, such as a crime scene, workplace, or home, is considered to originate from a single source (i.e., the geographical location itself). 
 	\item The composition of dust samples recovered on trace objects may be influenced by more than one source. 
 	\item The evidentiary object has visited at most one unknown geographical location in addition to a set of known locations (i.e, $Q \leq 1$). 
 \end{enumerate}
 
  Assumptions (a) and (b) are identical to the assumptions made to develop the original model proposed by \cite{Blei:2003}. Assumption (a) is reasonable since particles are not organised in any particular order in a dust sample. In practice, an appropriate sampling procedure will ensure that assumption (b) holds. Assumption (c) considers that the dust profile of any given dust source is characterised by the ``dust output'' of that source, and thus accounts for any prior cross-contamination between sites. This assumption may not be appropriate and may be investigated in future work. Assumptions (d) and (e) are critical for the inferences we want to make with this model: they allow us to make inferences on the origin of the dust recovered from trace objects of forensic interest in terms of geographical sampling locations from mixtures of dust sources. Finally, assumption (f) is made in light of the foundational nature of our work and currently constrains the number of geographical locations that can contribute to the mixture of dust in a trace sample. It is supported by recent work on the rates of loss and replacement of very small particles on the contact surface of footwear \citep{Stoney:2018}. Assumption (f) will be removed in future work.

\subsection{Defining dust samples}
We describe the generative process of a dust sample as follows. 
Notation is summarised in tables~\ref{table:notation1} and~\ref{table:notation2} in Appendix~\ref{appendix:notation}. The top part of figure~\ref{DAG} provides a graphical representation of this process. 
\begin{figure}[h]
	\vspace{8mm}
	\centering
	\includegraphics[width=9cm]{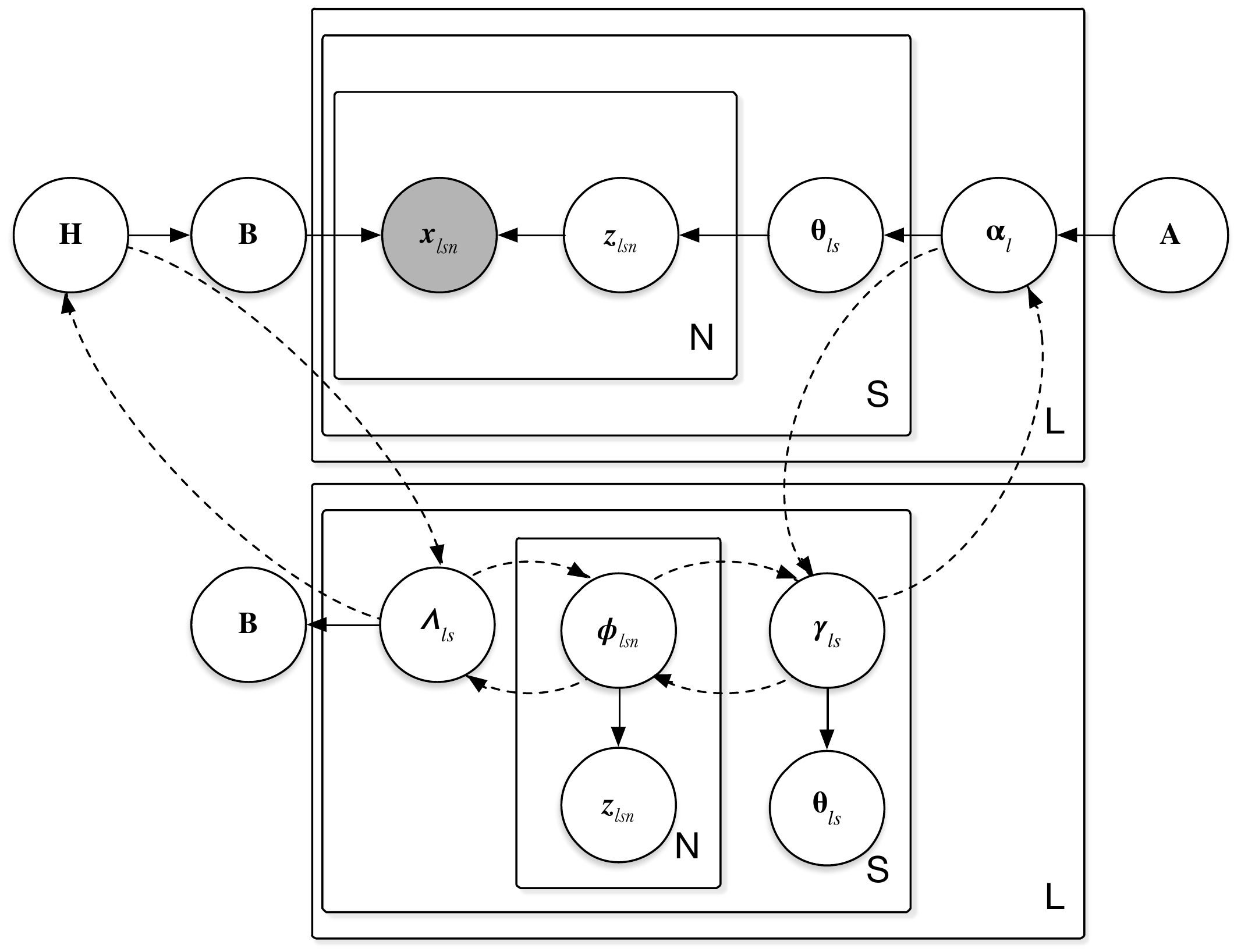} \\
	\caption{Generative process for a sample of dust particles (top part) and update process for the Variational Bayesian Inference algorithm (bottom part).}
	\label{DAG}
\end{figure}
\begin{enumerate}[(a)]
	\item Choose an $M \times T$ matrix $\mathbf{H}$ to represent the relative contributions of the different particle types to the dust profiles of each of the $M=Q+K$ sources that have the potential to contribute to the mixture. The $m$-th row of $\mathbf{H}$ corresponds to the parameters of a $T-$Dirichlet distribution that drives the mixing proportions of the $T$ particle types that characterise source $m$. 
	\item Choose an $L \times M$ matrix $\mathbf{A}$ to represent the relative contributions of the $M$ different sources to each of the $L$ locations from which we obtain dust samples. The $l$-th row, $\boldsymbol{\alpha}_l$, of $\mathbf{A}$ corresponds to the parameters of an $M-$Dirichlet distribution that drives the mixing proportions of the $M$ sources in samples obtained from location $l$. 
	\item For a set of dust samples obtained from known locations and on evidentiary objects, sample an $M \times T$ matrix $\mathbf{B}$  from  $\mathbf{H}$ to obtain the mixing proportions of the types of dust particles for each source of dust $m \in \{1, \dots, M\}$.	
	\item For a sample $\mathbf{x}_{ls}$ taken from location $l$:
	\begin{enumerate}[(i)]
		\item Sample $\boldsymbol{\mathbf{\theta}}_{ls} \sim$ Dirichlet$\left(\boldsymbol{\mathbf{\alpha}}_l\right)$ to obtain a vector of mixing proportions of the dust sources for the samples obtained from location $l$. 
		\item For each of the $N_{ls}$ particles, $\mathbf{x}_{lsn}$, in sample $\mathbf{x}_{ls}$: 
		\begin{enumerate}[(a)]
			\item Sample a source of dust $\mathbf{z}_{lsn} \sim$ Multinomial $\left(1, \boldsymbol{\mathbf{\theta}}_{ls}\right)$.  
			\item Sample a particle $\mathbf{x}_{lsn}  \sim$ Multinomial $\left(1, \boldsymbol{\mathbf{\beta}}_{{z_{lsn}}}\right)$, where $\boldsymbol{\mathbf{\beta}}_{{z_{lsn}}}$ represents the row of matrix  $\mathbf{B}$ for the source defined by $\mathbf{z}_{lsn}$. 
		\end{enumerate}
	\end{enumerate}
\end{enumerate}
We see that the model makes no assumptions pertaining to any sort of ordering or grouping of the particles or the locations in a dust sample. This is synonymous to the \textit{bag of words assumption} (i.e., exchangeability (\cite{Robert:2007}, page 159) that is commonly associated with topic modelling. 

This process can be represented by means of the Directed Acyclic Graph (DAG) depicted in figure~\ref{DAG} (top part). By making use of the DAG and the generative process described above, the probability of a sample of dust particles is given by:
\begingroup\makeatletter\def\f@size{8.0}\check@mathfonts
\begin{eqnarray}
	p\left(\mathbf{x}_{ls}|\boldsymbol{\mathbf{\alpha}}_l, \mathbf{H}\right)=\int \int p\left(\boldsymbol{\mathbf{\theta}}_{ls}|\boldsymbol{\mathbf{\alpha}}_l\right) p\left(\mathbf{B}|\mathbf{H}\right) \prod_{n=1}^N \sum_{\mathbf{z}_{lsn}} p\left(\mathbf{z}_{lsn}|\boldsymbol{\mathbf{\theta}}_{ls}\right)p\left(\mathbf{x}_{lsn}|\mathbf{z}_{lsn},\mathbf{B}\right)d\boldsymbol{\mathbf{\theta}}_{ls} d\mathbf{B}. 
\end{eqnarray}
The joint probability of a set of samples collected across multiple locations is then given by: 
\begin{eqnarray}
	p\left(\mathbf{X}|\mathbf{A}, \mathbf{H}\right)=\prod_{l=1}^L \prod_{s=1}^{S_l} \left[\int \int p\left(\boldsymbol{\mathbf{\theta}}_{ls}|\boldsymbol{\mathbf{\alpha}}_l\right) p\left(\mathbf{B}|\mathbf{H}\right) \prod_{n=1}^N \sum_{\mathbf{z}_{lsn}} p\left(\mathbf{z}_{lsn}|\boldsymbol{\mathbf{\theta}}_{ls}\right)p\left(\mathbf{x}_{lsn}|\mathbf{z}_{lsn},\mathbf{B}\right)d\boldsymbol{\mathbf{\theta}}_{ls} d\mathbf{B}\right] \label{joint_prob}. 
\end{eqnarray}
The distributions represented by each node in the top part of figure~\ref{DAG} are given by:
\begin{eqnarray}
	p\left(\boldsymbol{\mathbf{\theta}}_{ls}|\boldsymbol{\mathbf{\alpha}}_l\right)&=&\frac{\Gamma\left(\sum_{j=1}^M \alpha_{lj}\right)}{\prod_{m=1}^M \Gamma\left(\alpha_{lm}\right)}\prod_{m=1}^M\theta_{lsm}^{\alpha_{lm}-1} \label{theta.dist} \\
	p\left(\mathbf{B}|\mathbf{H}\right)&=&\prod_{m=1}^M\frac{\Gamma\left(\sum_{j=1}^T \eta_{mj}\right)}{\prod_{t=1}^T\Gamma\left(\eta_{mt}\right)}\prod_{t=1}^T\beta_{mt}^{\eta_{mt}-1} \label{b.dist} \\
	p\left(\mathbf{z}_{sn}|\boldsymbol{\mathbf{\theta}}_{ls}\right)&=&\prod_{m=1}^M \theta_{lsm}^{z_{lsnm}} \label{z.dist} \\
	p\left(\mathbf{x}_{sn}|\mathbf{z}_{sn}, \mathbf{B}\right) &=& \prod_{m=1}^M \prod_{t=1}^T \beta_{mt}^{x_{lsnt} z_{lsnm}} \label{x.dist}
\end{eqnarray}
\section{Assigning the model parameters}
We use Variational Bayesian Inference (VBI) to assign the \normalsize \textit{approximate posterior distribution} $p\left(\mathbf{\Theta}, \mathbf{B}|\mathbf{X}\right)$ of the model's parameters given a set of exchangeable observations $\mathbf{X}$ obtained from $L$ geographical locations and trace objects. This is achieved by maximising the lower bound function defined by the negative Kullback-Leibler divergence of the joint distribution  $p\left(\mathbf{X}, \mathbf{\Theta}, \mathbf{B}, \mathbf{Z}\right)$, and the variational distribution $q\left(\mathbf{\Theta}, \mathbf{B}, \mathbf{Z}\right)$ (\cite{Bishop:2006}, Chapter 10). We introduce the variational parameters, $\mathbf{\Gamma}, \mathbf{\Lambda}$, and $\mathbf{\Phi}$ to break the dependencies that exist between $\mathbf{\Theta}, \mathbf{B}$, and $\mathbf{Z}$ (see figure~\ref{DAG} (bottom part)), and we define $q\left(\mathbf{\Theta}, \mathbf{B}, \mathbf{Z}\right)$ as: 
\begingroup\makeatletter\def\f@size{8.0}\check@mathfonts
\begin{eqnarray}
	q\left(\mathbf{\Theta}, \mathbf{B}, \mathbf{Z}|\mathbf{\Gamma}, \mathbf{\Lambda}, \mathbf{\Phi}\right) = \prod_{l=1}^L \prod_{s=1}^S \left[q_{ls}\left(\boldsymbol{\mathbf{\theta}}_{ls}| \boldsymbol{\mathbf{\gamma}}_{ls}\right)\prod_{n=1}^N q_{ls}\left(\mathbf{z}_{lsn}|\boldsymbol{\mathbf{\phi}}_{lsn}\right)\prod_{m=1}^M q_m\left(\boldsymbol{\mathbf{\beta}}_{m}|\boldsymbol{\mathbf{\lambda}}_{lsm}\right)\right] \label{var_dist}.
\end{eqnarray}
The ``E-Step'' of our implementation of VBI maximises the lower bound function with respect to each of the variational parameters, \normalsize $\mathbf{\Gamma}$, $\mathbf{\Lambda}$, and $\mathbf{\Phi}$, while maintaining fixed values of $\mathbf{\Theta}, \mathbf{B}$, and $\mathbf{Z}$; the ``M-Step'' maximises the lower bound function with respect to the global latent parameters $\mathbf{H}$ and $\mathbf{A}$, while keeping the variational parameters obtained in the E-step fixed. Each step is itself an iterative process that repeats until some convergence criteria is satisfied.  From equation~(\ref{var_dist}), we note that $\boldsymbol{\mathbf{\gamma}}_{ls}, \mathbf{\Lambda}_{ls}$, and $\mathbf{\Phi}_{ls}$ can be updated independently to minimise the KL divergence between $q\left(\mathbf{\Theta}, \mathbf{B}, \mathbf{Z}|\mathbf{\Gamma}, \mathbf{\Lambda}, \mathbf{\Phi}\right)$ and $p\left(\mathbf{X},\mathbf{\Theta}, \mathbf{B}, \mathbf{Z}\right)$ for each sample. For ease of notation, we now suppress the explicit conditioning on the variational parameters, and shorten $q\left(\mathbf{\Theta}, \mathbf{B}, \mathbf{Z}|\mathbf{\Gamma}, \mathbf{\Lambda}, \mathbf{\Phi}\right)$ to $q\left(\mathbf{\Theta}, \mathbf{B}, \mathbf{Z}\right)$.
\subsection{The lower bound function for mixtures of dust particles}
The lower bound function mentioned above is the sum of expectations of each of the latent parameters, taken with respect to the variational distribution, $q\left(\mathbf{\Theta}, \mathbf{B}, \mathbf{Z}\right)$, as shown in equation~(\ref{lb_exp}). Note that each line in the last equality of equation~(\ref{lb_exp}) corresponds to one expectation. In addition, we note that $\Psi$ is the digamma function, which gives the logarithmic derivative of the Gamma function, such that $\Psi\left(\cdot\right) = \frac{d\log\Gamma\left(\cdot\right)}{d\cdot}$.  
\footnotesize
\begin{eqnarray}
	\mathscr{L}\left(q\left(\mathbf{\Theta}, \mathbf{B}, \mathbf{Z}\right)\right) &=& \int_{\mathbf{\Theta}}\int_{\mathbf{B}}\int_{\mathbf{Z}} q\left(\mathbf{\Theta}, \mathbf{B}, \mathbf{Z}\right)\log p\left(\mathbf{X},\mathbf{\Theta},\mathbf{B}, \mathbf{Z}\right) d\mathbf{Z}d\mathbf{B}d\mathbf{\Theta}\hspace{2mm}-\nonumber \\
	&\quad& \int_{\mathbf{\Theta}}\int_{\mathbf{B}}\int_{\mathbf{Z}} q\left(\mathbf{\Theta}, \mathbf{B}, \mathbf{Z}\right) \log q\left(\mathbf{\Theta}, \mathbf{B}, \mathbf{Z}\right) d\mathbf{Z}d\mathbf{B}d\mathbf{\Theta} \nonumber \\
	&=& \mathbb{E}_q\left[\log p\left(\mathbf{X}|\mathbf{Z}, \mathbf{B}\right)\right]+\mathbb{E}_q\left[\log p\left(\mathbf{\Theta}|\mathbf{A}\right)\right]+\mathbb{E}_q\left[\log p\left(\mathbf{Z}|\mathbf{\Theta}\right)\right] +  \nonumber \\
	&\quad& \mathbb{E}_q\left[\log p\left(\mathbf{B}|\mathbf{H}\right)\right]-\mathbb{E}_q\left[\log q\left(\mathbf{\Theta}\right)\right]-\mathbb{E}_q\left[\log q\left(\mathbf{Z}\right)\right]-\mathbb{E}_q\left[\log q\left(\mathbf{B}\right)\right]  \label{lb_exp}
\end{eqnarray}
The\normalsize first four expectations are developed using equations~(\ref{theta.dist}) - (\ref{x.dist}) above:
\begingroup\makeatletter\def\f@size{8.0}\check@mathfonts
\begin{eqnarray}
	\mathbb{E}_q\left[\log p\left(\mathbf{X}|\mathbf{Z}, \mathbf{B}\right)\right] &=& \sum_{l=1}^L \sum_{s=1}^S \sum_{n=1}^N \sum_{m=1}^M \sum_{t=1}^T x_{lsnt} \phi_{lsnm} \left(\Psi\left(\lambda_{lsmt}\right)-\Psi\left(\sum_{j=1}^T \lambda_{lsmj}\right)\right) \nonumber \\
	\mathbb{E}_q\left[\log p\left(\mathbf{\Theta}|\mathbf{A}\right)\right] &=& \sum_{l=1}^L \sum_{s=1}^S \sum_{n=1}^N \sum_{m=1}^M \phi_{lsnm}\left(\Psi\left(\gamma_{lsm}\right)-\Psi\left(\sum_{j=1}^M \gamma_{lsj}\right)\right) \nonumber  \\
	\mathbb{E}_q\left[\log p\left(\mathbf{Z}|\mathbf{\Theta}\right)\right] &=& \sum_{l=1}^L \sum_{s=1}^S \left[\log \Gamma\left(\sum_{j=1}^M \alpha_{lj}\right)-\sum_{m=1}^M \log \Gamma\left(\alpha_{lm}\right)+\sum_{m=1}^M \left(\alpha_{lm}-1\right)\left(\Psi\left(\gamma_{lsm}\right)-\Psi\left(\sum_{j=1}^M \gamma_{lsj}\right)\right)\right] \nonumber 
\end{eqnarray}
\begin{eqnarray}
	\mathbb{E}_q\left[\log p\left(\mathbf{B}|\mathbf{H}\right)\right] &=& \sum_{l=1}^L \sum_{s=1}^S \sum_{m=1}^M \left[\log \Gamma\left(\sum_{j=1}^T \eta_{mj}\right) - \sum_{t=1}^T \log \Gamma\left(\eta_{mt}\right)+\sum_{t=1}^T\left(\eta_{mt}-1\right)\left(\Psi\left(\lambda_{lsmt}\right)-\Psi\left(\sum_{j=1}^T \lambda_{lsmj}\right)\right)\right]. \nonumber
\end{eqnarray}
The\normalsize last three expectations are developed using the entropies of the distributions corresponding to each of the latent parameters, given by equations~(\ref{theta.dist}) - (\ref{z.dist}):
\begingroup\makeatletter\def\f@size{8.0}\check@mathfonts
\begin{eqnarray}
	-\mathbb{E}_q\left[\log q\left(\mathbf{\Theta}\right)\right] &=& \sum_{l=1}^L \sum_{s=1}^S \left[ -\log \Gamma\left(\sum_{j=1}^M \gamma_{lsj}\right) + \sum_{m=1}^M \log \Gamma\left(\gamma_{lsm}\right)-\sum_{m=1}^M\left(\gamma_{lsm}-1\right)\left(\Psi\left(\gamma_{lsm}\right)-\Psi\left(\sum_{j=1}^M \gamma_{lsj}\right)\right)\right] \nonumber \\
	-\mathbb{E}_q\left[\log q\left(\mathbf{Z}\right)\right] &=& \sum_{l=1}^L \sum_{s=1}^S \sum_{m=1}^M \left[-\log \Gamma\left(\sum_{j=1}^T \lambda_{slmj}\right)+\sum_{t=1}^T \log \Gamma\left(\lambda_{slmt}\right) - \sum_{t=1}^T\left(\lambda_{slmt}-1\right)\left(\Psi\left(\lambda_{slmt}\right)-\Psi\left(\sum_{j=1}^T \lambda_{slmj}\right)\right)\right] \nonumber \\
	-\mathbb{E}_q\left[\log q\left(\mathbf{B}\right)\right] &=& - \sum_{l=1}^L \sum_{s=1}^S \sum_{n=1}^N \sum_{m=1}^M \phi_{lsnm} \log \phi_{lsnm}. \nonumber
\end{eqnarray}
By maximising equation~(\ref{lb_exp}) with respect to each of the variational parameters, we obtain update equations for \normalsize $\phi_{lsnm}, \gamma_{lsm}$, and $\lambda_{lsmt}$:
\begin{eqnarray}
	&\phi_{lsnm} \propto \exp\left\{\sum_{t=1}^T x_{lsnt}\left(\Psi\left(\lambda_{lsmt}\right)-\Psi\left(\sum_{j=1}^T \lambda_{lsmj}\right)\right)+\Psi\left(\gamma_{lsm}\right)\right\}&  \label{phi_update} \\
	&\gamma_{lsm} = \alpha_{lm}+\sum_{n=1}^N \phi_{lsnm}\label{gamma_update}& \\
	&\lambda_{lsmt} = \eta_{mt}+ \sum_{n=1}^N x_{lst}\phi_{lsnm}& \label{lambda_update}
\end{eqnarray}
These updates for the variational parameters are used in the E-Step of the VBI algorithm described in algorithm~\ref{EstepAlg}. Assigning values to the global parameters \normalsize $\mathbf{H}$ and $\mathbf{A}$ in the M-Step requires using optimisation techniques, since tractable maximum likelihood solutions do not exist. We use the L-BFGS-B method \citep{Byrd:1994, Zhu:1994} to obtain the matrices $\mathbf{H}$ and $\mathbf{A}$. For more information pertaining to the L-BFGS-B method and the M-Step, see Appendix~\ref{appendix:MStep}.
\begin{algorithm}[h]
	\caption{E-Step - Updating the variational parameters \label {EstepAlg}}	
	\For {each sample}{
		Initialise $\mathbf{\Gamma}$:=$\mathbf{A}+\left(\sum_{n=1}^N x_{lsnt}\right)/\left(M*\sum_{l=1}^L S_l\right)$; \\
		Initialise $\mathbf{\Phi}_{ls}$ := $1/M$; \\
		Initialise $\mathbf{\Lambda}_{ls}$ := $\mathbf{H}$; \\
		\While{$\mathscr{L}\left(q\left(\mathbf{\Theta}, \mathbf{B}, \mathbf{Z}\right)\right)$ has not converged}{
				$\phi_{lsnm} \propto \exp\left\{\sum_{t=1}^T x_{lsnt}\left(\Psi\left(\lambda_{lsmt}\right)-\Psi\left(\sum_{j=1}^T \lambda_{lsmj}\right)\right)+\Psi\left(\gamma_{lsm}\right) \right\}$; \label{phi_update} \\
			 	$\gamma_{lsm} = N^{-1}\left(\alpha_{lm}+\sum_{n=1}^N \phi_{lsnm}\right)$; \label{gamma_update} \\
			 	$\lambda_{lsmt} = N^{-1}\left(\eta_{mt}+ \sum_{n=1}^N x_{lst}\phi_{lsnm}\right)$ ;\label{lambda_update} \\ 
		}
	}			
\end{algorithm}
\subsection{Initialisation of the model}
By assumption (d), our model considers that the dust observed at any given geographical location originates from a single source. Hence, the strategy behind the proposed model is to learn the dust profile of the $K$ known sources by obtaining ``pure'' samples from the corresponding locations, and to infer the profile of the remaining $Q$ unknown sources (corresponding to $Q$ locations that cannot be studied) through the deconvolution of the samples recovered on the surface of the trace objects. This process also provides information on the respective contributions of the $M=Q+K$ sources to the mixtures observed on the trace objects. 

To ensure that each geographical location (known and unknown) is uniquely associated with a single source, we constrain, $\mathbf{A}$, the $L\times M$ matrix of Dirichlet parameters controlling the mixing proportions of dust sources at each location. Each column of $\mathbf{A}$ corresponds to one of the $M$ sources that may have contributed dust to the different locations, and $K$ of the $L$ rows of $\mathbf{A}$ correspond to known locations where ``pure'' samples were collected. Constraining $\mathbf{A}$ requires to heavily weigh the $m$-th element of the $l$-th row of $\mathbf{A}$, when $m=l$ and $m,l \leq K$. For rows of $\mathbf{A}$ corresponding to the evidentiary samples, we use a flat $M-$Dirichlet distribution to reflect that, before running the algorithm, we consider that all dust sources are equally likely to have contributed to these samples. The values of the rows associated with the $K$ known locations are not updated by the algorithm, while the rows associated with the evidentiary samples are updated during the ``M-step'' of the algorithm. There is currently a level of arbitrariness involved in the selection of the weight values for the known locations. The choice of weights appears to impact the convergence of the optimisation of $\mathbf{A}$. More objective ways of selecting these weights or more robust optimisation methods need to be investigated.

Matrix $\mathbf{H}$ is initialised with flat Dirichlet distributions for all rows. All rows are updated during the M-step of the algorithm to learn the dust profiles of the different sources potentially contributing to the evidentiary samples. 
\section{Inferences on sources' dust profiles and mixing proportions}
\label{PostInf}
Following convergence of the algorithm, we obtain updated Dirichlet distribution parameter matrices $\mathbf{A}$ and $\mathbf{H}$. The marginal distribution of each of the Dirichlet distributions in the rows of $\mathbf{A}$ and $\mathbf{H}$ gives the posterior distributions for the multinomial parameters $\mathbf{\Theta}_{ls}$ and $\mathbf{B}$, respectively. Hence, 
\begin{enumerate}[(a)]
	\item the contribution of the $m$-th source to the $s$-th sample from location $l$ is~$\theta_{lsm}\sim$ Beta$\left(\alpha_{lm}, \sum_{i\neq m} \alpha_{li}\right)$. Note that, by construction, the expectation of $\theta_{lsm}$ will be very close to 1 if location $l$ is one of the $K$ known locations and $l=m$.
	\item the proportion of the $t$-th particle type in the $m$-th source is \ $\beta_{mt}\sim$ Beta$\left(\eta_{mt}, \sum_{i\neq t} \eta_{mi}\right)$. 
\end{enumerate}
\section{Worked example}
\label{worked_example}
As an example, the algorithm presented above is used to resolve two mixtures of dust particles provided by Stoney Forensic, Inc. (Chantilly, VA, USA) under different scenarios. The data set is composed of ``pure'' and mixed samples of dust from two locations, labeled AT and LQ. These locations are extensively described in \cite{Stoney:2018}. The control and trace materials consist in (1) twelve samples of dust knowingly obtained from each of the two locations, and (2) two trace samples consisting in mixtures of dust obtained by mixing known proportions of dust from the two known locations. A dust sample is characterised by a vector of counts for fourteen particle types. The data set is summarised in tables~\ref{ATsamps}, \ref{LQsamps} and~\ref{mixsamps}.
Our model is used to resolve the dust mixtures in the trace samples presented in table~\ref{mixsamps} under three different scenarios: 
\begin{enumerate}[(a)]
\item In the first scenario, both sources are considered known and can be sampled from ($K=2,Q=0$);
\item In the second scenario, location AT is known and can be sampled from, while LQ is considered unknown and cannot be studied ($K=1,Q=1$);
\item In the last scenario, location LQ is known and can be sampled from, while AT is considered unknown and cannot be studied ($K=1,Q=1$).
\end{enumerate}
\renewcommand{\arraystretch}{0.35}
\begin{table}[H]
	\centering
	\caption{Twelve samples obtained from location AT. 	\label{ATsamps}}
	\renewcommand{\arraystretch}{0.9}
	\resizebox{15cm}{!}{
	\begin{tabular}{p{2.9 cm} p{0.5 cm} p{0.5 cm} p{0.5 cm} p{0.5 cm} p{0.5 cm}p{0.5 cm}p{0.5 cm}p{0.5 cm}p{0.5 cm}p{0.5 cm}p{0.5 cm}p{0.5 cm}}	
	\hline
	\em{}&{1} &{2}&{3}&{4}&{5}&{6}&{7}&{8}&{9}&{10}&{11}&{12} \\
	\hline\hline
	Alkali Feldspar&189&182&200&184&254&182&181&178&139&193&229&204 \\
	Alterite	&21&20&9&20&15&11&21&19&12&28&32&20 \\
	Biotite&1&4&4&1&3&0&1&3&0&1&0&0 \\
	Epidote&3&3&7&6&11&12&3&7&12&5&4&7 \\
	High Index&3&2&1&7&3&3&2&2&3&2&1&4 \\
	Hornblende&0&2&2&4&5&4&2&3&2&1&0&2 \\
	Iron Oxides&9&4&1&5&5&7&7&9&5&6&7&6 \\
	Lithic Fragments&3&0&7&12&3&2&3&2&3&3&4&0 \\
	Muscovite&0&0&1&1&0&3&0&0&0&3&0&0 \\
	Opaques&16&14&10&9	&37&18&25&20&42&16&9&15 \\
	Plagioclase&5&0&2&5&7&1&0&2&0&10&5&10 \\
	Quartz&74&74&75&63&112&90&62&71&101&56&54&94 \\
	Titanite&0&0&0&0&0&1&1&0&0&0&0&0 \\
	Other&11&11&5&12&17&23&3&6&8&9&9&8 \\
	\hline
	\end{tabular}}
\end{table} 
\begin{table}[H]
	\centering
	\caption{Twelve samples obtained from location LQ. 	\label{LQsamps}}
	\renewcommand{\arraystretch}{0.9}
	\resizebox{15cm}{!}{
	\begin{tabular}{p{2.9 cm} p{0.5 cm} p{0.5 cm} p{0.5 cm} p{0.5 cm} p{0.5 cm}p{0.5 cm}p{0.5 cm}p{0.5 cm}p{0.5 cm}p{0.5 cm}p{0.5 cm}p{0.5 cm}}	
	\hline
	\em{}&{1} &{2}&{3}&{4}&{5}&{6}&{7}&{8}&{9}&{10}&{11}&{12} \\
	\hline\hline
	Alkali Feldspar&18&26&29&20&31&33&28&30&39&22&30&22 \\
	Alterite&4&4&4&6&7&7&10&4&5&12&9&10 \\
	Biotite&16&10&22&13&10&12&26&13&11&25&18&20 \\
	Epidote&10&5&13&11&7&7&8&19&6&11&13&5 \\
	High Index&3&0&	1&2&2&0&0&1&0&0&	2&2 \\
	Hornblende&73&	55&	64&	68&	61&	91&	93&	68&	51&	73&	82&	75 \\
	Iron Oxides&0&	0&0&0&0&1&0&0&0&2&0&	0 \\
	Lithic Fragments&5&7&6&6&2&4&0&6&3&6&7&11 \\
	Muscovite&0&3&0&2&2&0&1&4&0&2&5&1 \\
	Opaques&5&0&2&4&8&10&8&5&3&7&4&3 \\
	Plagioclase&46&37&47&45&52&39&14&16&13&33&35&27 \\
	Quartz&153&159&151&145&174&150&128&161&195&134&137&156 \\
	Titanite&2&4&6&1&4&6&5&6&2&2&4&2 \\
	Other&3&5&1&4&9&6&8&2&3&5&2&5 \\
	\hline 
	\end{tabular}}
\end{table} 
\begin{table}[H]
	\centering
	\tabcolsep=0.06cm
	\caption{Two samples obtained by mixing dust from AT and LQ in known proportions. The proportions are indicated in the last column.	\label{mixsamps}}	
	\renewcommand{\arraystretch}{0.9}
	\resizebox{\textwidth}{!}{
	\begin{tabular}{p{1.5 cm} p{0.8cm} p{0.8cm} p{0.8cm} p{0.8cm} p{0.8cm} p{0.8cm} p{0.8cm} p{0.8cm} p{0.8cm} p{0.8cm} p{0.8cm} p{0.8cm} p{0.8cm} p{0.8cm} p{1.6cm}}
	\hline
	\em{}&{1}&{2}&{3}&{4}&{5}&{6}&{7}&{8}&{9}&{10}&{11}&{12}&{13}&{14}&{LQ/AT} \\ 
	\hline\hline
	{Trace 1}&312&31&5&12&5&16&9&7&1&32&12&151&1&17&0.10/0.90 \\
	{Trace 2}&104&16&23&16&2&100&2&9&3&13&48&240&5&10&0.80/0.20\\
	\hline
	\end{tabular}}
\end{table} 
\renewcommand{\arraystretch}{1}	
\subsection{Resolving the mixtures in table~\ref{mixsamps} when both locations are known}
\label{bothknown_experiment}
In this example, the algorithm is provided with all 26 samples described above: 12 ``pure'' control samples from each location and 2 ``mixed'' trace samples composed of locations AT and LQ in the proportions specified in the last column of table \ref{mixsamps}. This scenario serves primarily to demonstrate the effectiveness of the algorithm when all locations in a mixture can be observed. We initialise the algorithm by setting the matrices $\mathbf{H}$ and $\mathbf{A}$ to some initial values. To allow the model to freely determine the particle profiles of the two sources in the dust mixtures, the matrix $\mathbf{H}$ is set to:
\setcounter{MaxMatrixCols}{20}
\[
\scalemath{0.85}{
	\mathbf{H}_{initial}=\begin{bmatrix}
		1 & 1 & 1 &  1 & 1 & 1 & 1 & 1 & 1 & 1 & 1 &  1 & 1 & 1	\\
		1 & 1 & 1 &  1 & 1 & 1 & 1 & 1 & 1 & 1 & 1 &  1 & 1 & 1	\\
	\end{bmatrix}		
	}.
\]
The first row of this matrix corresponds to the relative contributions of each of the fourteen particle types to source AT, while the second row of this matrix corresponds to the relative contributions of each of the fourteen particle types to source LQ.

To ensure that the algorithm correctly learns the dust profiles of the two known sources from the samples obtained from \textit{locations} AT and LQ, the rows of matrix $\mathbf{A}$ associated with these samples are heavily weighted in the dimension corresponding to \textit{sources} AT and LQ, while the rows of matrix $\mathbf{A}$ corresponding to the trace samples are set to 1:
\[
\scalemath{0.75}{
	\mathbf{A}_{initial} = \begin{bmatrix}
		\boldsymbol{\mathbf{\alpha}}_{AT,1} \\
		\vdots \\
		\boldsymbol{\mathbf{\alpha}}_{AT,12} \\
		\boldsymbol{\mathbf{\alpha}}_{LQ,1} \\
		\vdots \\
		\boldsymbol{\mathbf{\alpha}}_{LQ,12} \\
		\boldsymbol{\mathbf{\alpha}}_{e_1,1} \\
		\boldsymbol{\mathbf{\alpha}}_{e_2,1} \\
	\end{bmatrix}
	= \begin{bmatrix} 
	150 & 1 \\
	\vdots & \vdots \\
	150 & 1 \\
	1 & 150 \\
	\vdots & \vdots \\
	1 & 150 \\
	1 & 1  \\
	1 & 1  \\
	\end{bmatrix}
}.
\]
The first column of this matrix corresponds to the relative contribution of source AT to the dust sample being considered. Likewise, the second column of this matrix corresponds to the relative contribution of source LQ to the dust sample being considered.

Upon introducing the samples representing the known sources and the two trace samples into the model and observing convergence, we obtain:
\setcounter{MaxMatrixCols}{20}
\[
\scalemath{0.70}{
	\mathbf{H}_{converged}=\begin{bmatrix}
	2315.42 & 227.43 & 17.50 & 79.56 & 33.14 & 25.33 & 70.77 & 41.80 & 7.76 & 230.05 & 45.80 & 922.44 & 1.98 & 121.65 \\
	323.67 & 81.57 & 195.91 & 115.05 & 12.87 & 854.48 & 2.82 & 63.16 & 20.07 & 58.44 & 403.62 & 1843.22 & 44.05 & 52.90 \\
	\end{bmatrix},		
	}
\]
\[
\scalemath{0.75}{
	\mathbf{A}_{converged} = \begin{bmatrix}
		\boldsymbol{\mathbf{\alpha}}_{e_1,1} \\
		\boldsymbol{\mathbf{\alpha}}_{e_2,1} \\
	\end{bmatrix}
	= \begin{bmatrix} 
		8978.75 & 1021.15 \\
		2007.53 & 7992.69 \\
	\end{bmatrix}
}.
\]
Note that, by design of the algorithm, the rows corresponding to $\boldsymbol{\mathbf{\alpha}}_{AT,s}$ and $\boldsymbol{\mathbf{\alpha}}_{LQ,s}$ are not updated (and are therefore not represented in $\mathbf{A}_{converged}$). The rows of these matrices are the parameters of the posterior marginal distributions described in section~\ref{PostInf}. $\mathbf{H}_{converged}$ enables us to study the distributions of $\boldsymbol{\mathbf{\beta}}_{AT}$ and $\boldsymbol{\mathbf{\beta}}_{LQ}$, which represent the particle profiles of the sources present in the dust samples. The last two rows of $\mathbf{A}_{converged}$ enables us to study the distributions of $\boldsymbol{\mathbf{\theta}}_{e_1,1}$ and $\boldsymbol{\mathbf{\theta}}_{e_2,1}$, which represent the mixing proportions of the two sources in the evidentiary samples.  The resulting marginal posterior distributions of $\boldsymbol{\mathbf{\beta}}_{AT}$ and $\boldsymbol{\mathbf{\beta}}_{LQ}$ are displayed in figure~\ref{ATLQ_both_profiles}. The resulting marginal posterior distributions of $\boldsymbol{\mathbf{\theta}}_{e_1,1}$ and $\boldsymbol{\mathbf{\theta}}_{e_2,1}$ are displayed in figure~\ref{ATLQ_both_2traces}.

Figure~\ref{ATLQ_both_profiles} shows that the model is able to effectively extract the dust profiles of the sources. All posterior distributions are sharply centred on their mean and mode.

\begin{figure}[H]
	\centering
	\includegraphics[scale=0.30]{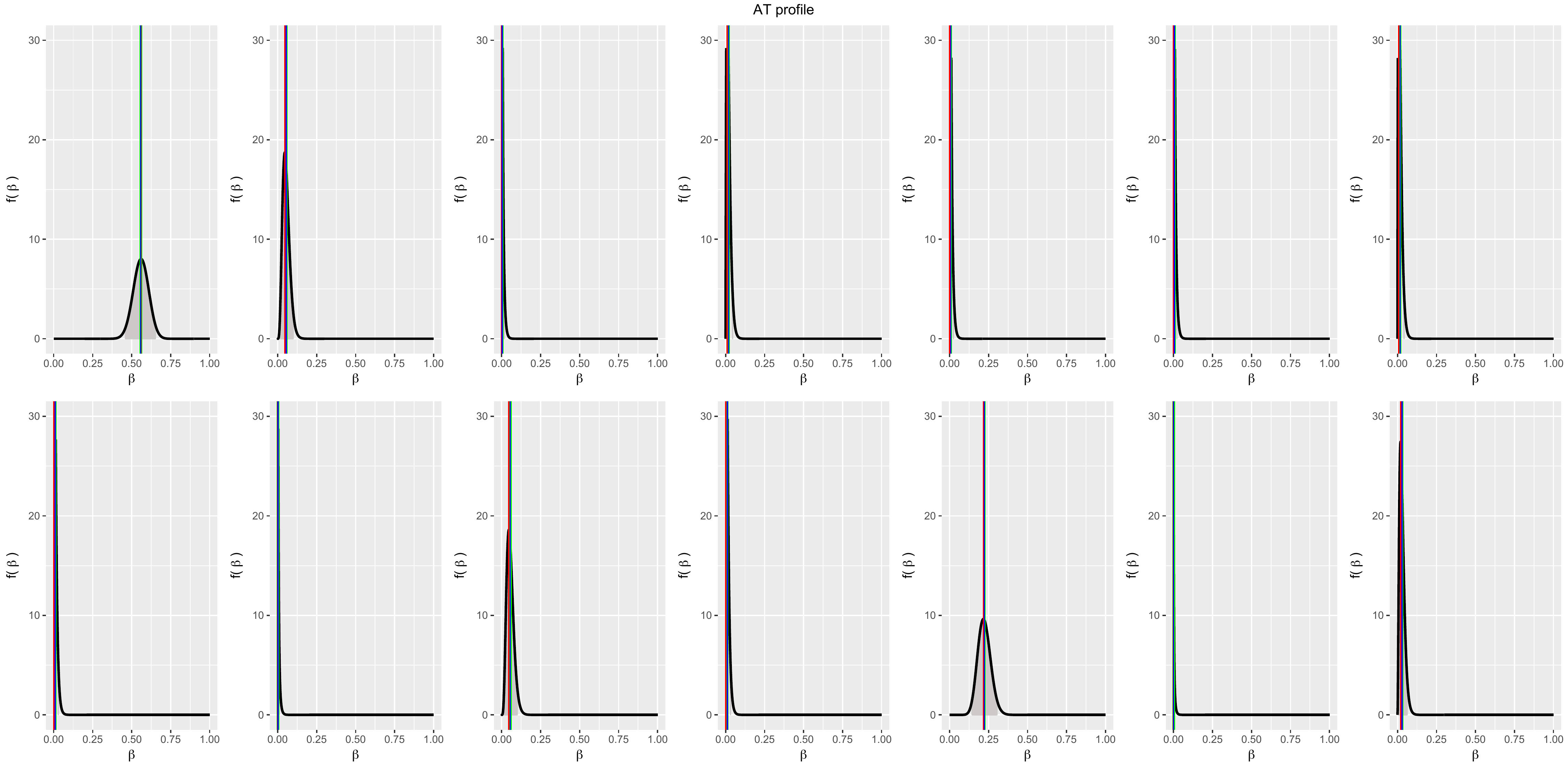} \\
	\vspace{1mm}
	\includegraphics[scale=0.30]{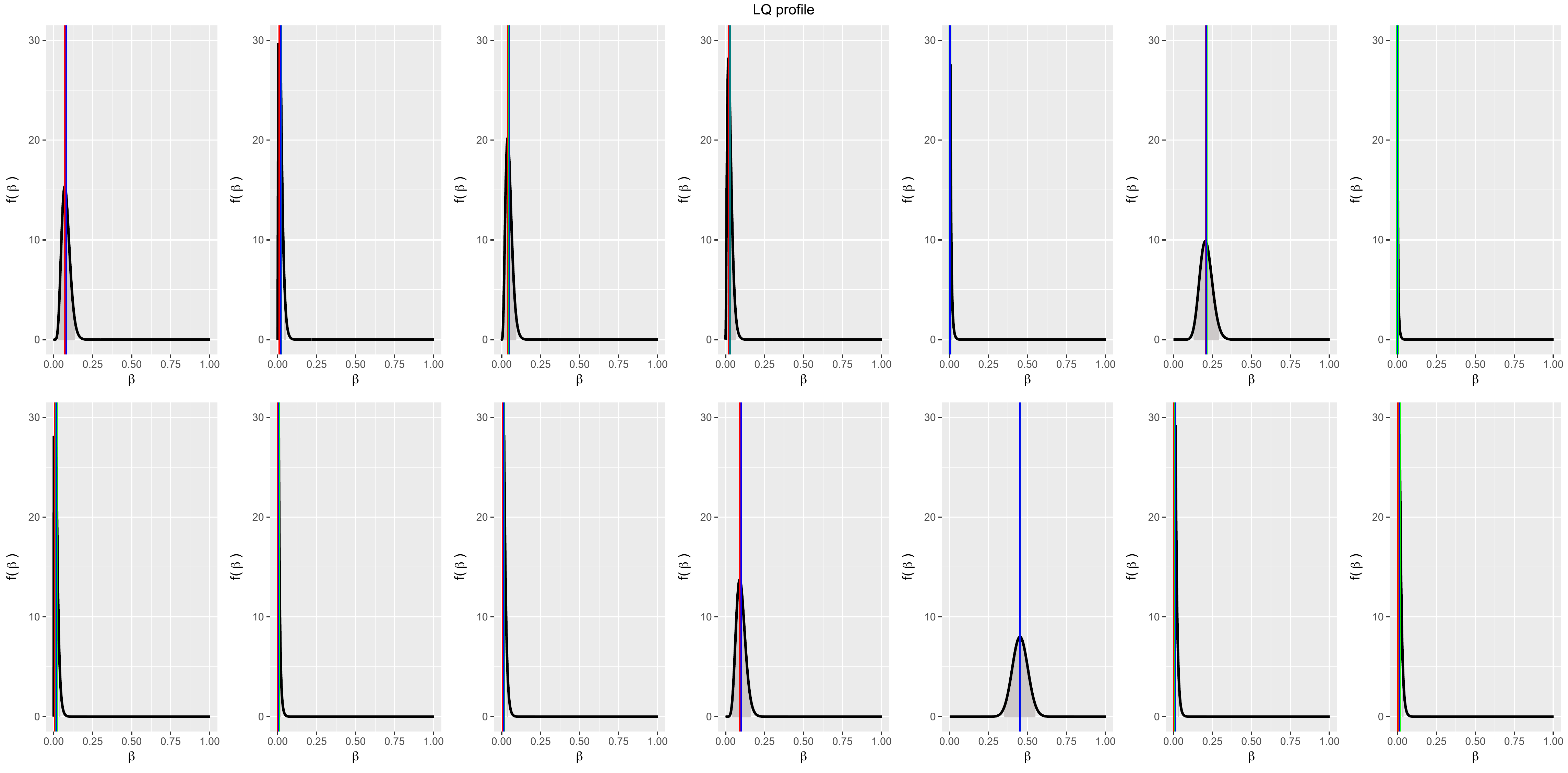}
	\caption{Plots of posterior distributions corresponding to the elements of $\mathbf{B}$ for location AT (top two rows) and location LQ (bottom two rows) when both locations are known.  Each plot is associated with one of the fourteen particle types. The vertical blue line corresponds to the point estimates of the mixing proportions, the vertical green line corresponds to the mean of the resulting posterior distribution, and the vertical red line corresponds to the mode of the resulting posterior distribution.The grey shaded region corresponds to the 95$\%$ HPDI.}
	\label{ATLQ_both_profiles}
\end{figure}

Figure~\ref{ATLQ_both_2traces} shows that the model is also able to extract the mixing proportions of the locations within the dust mixtures. The known mixing proportions are within the 95\% Highest Posterior Density Intervals (HPDI's), and the posterior distributions show little variance.

\begin{figure}[H]
	\centering
	\includegraphics[scale=0.25]{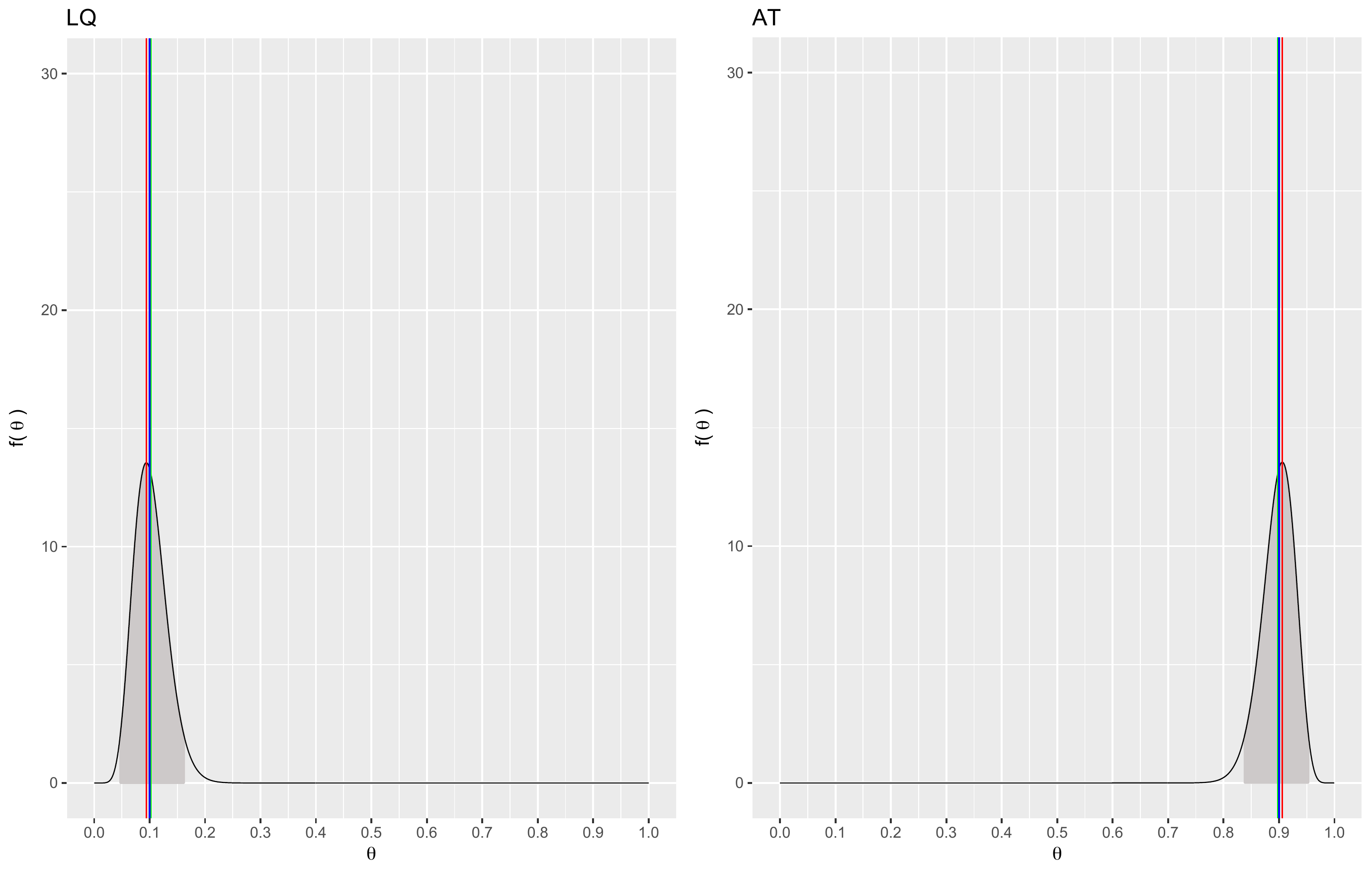} \hspace{2mm}
	\includegraphics[scale=0.25]{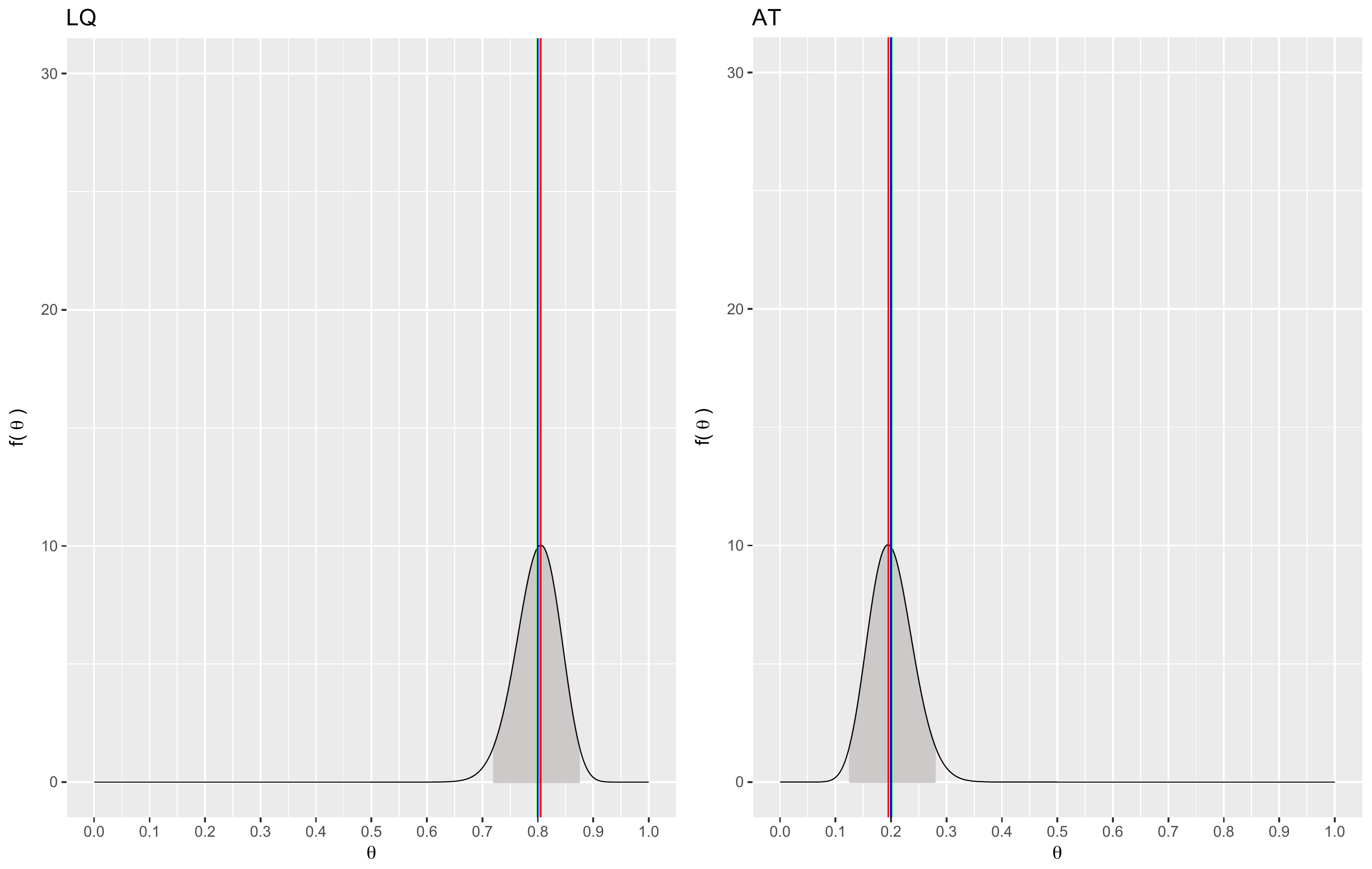} 
	\caption{Plots of posterior distributions corresponding to the elements of $\mathbf{A}$ when both AT and LQ are known and ''pure`` samples can be obtained from both. The first pair of plots on the left corresponds to a trace mixture where 90$\%$ of the particles originate from location AT and 10$\%$ of the particles originate from location LQ (first row of table \ref{mixsamps}). The second pair of plots on the right corresponds to a trace mixture where 20$\%$ of the particles originate from location AT, and 80$\%$ originate from location LQ  table (second row of table \ref{mixsamps}). The vertical blue line corresponds to the known mixing proportion, the vertical green line corresponds to the mean of the resulting posterior distribution, and the vertical red line corresponds to the mode of the resulting posterior distribution.The grey shaded region corresponds to the 95$\%$ HPDI.}
	\label{ATLQ_both_2traces}
\end{figure}

\subsection{Resolving the mixtures in table~\ref{mixsamps} when only location AT is known}
\label{ATKnown}

In this example, the algorithm is provided with 14 samples described above: 12 ``pure'' samples from location AT and 2 ``mixed'' samples. Matrix $\mathbf{H}$ is initialised as in the previous example, since we are still interested in learning the dust profiles of both sources. 

However, matrix $\mathbf{A}$ has a different number of rows to reflect that no sample representing source LQ has been observed. Hence, only the rows of matrix $\mathbf{A}$ associated with the samples obtained from location AT are heavily weighted in the dimension corresponding to source AT. As before, the rows of $\mathbf{A}$ corresponding to the trace samples are set to 1:
\[
\scalemath{0.75}{
	\mathbf{A}_{initial} = \begin{bmatrix}
		\boldsymbol{\mathbf{\alpha}}_{AT,1} \\
		\vdots \\
		\boldsymbol{\mathbf{\alpha}}_{AT,12} \\
		\boldsymbol{\mathbf{\alpha}}_{e_1,1} \\
		\boldsymbol{\mathbf{\alpha}}_{e_2,1} \\
	\end{bmatrix}
	= \begin{bmatrix} 
	150 & 1 \\
	\vdots & \vdots \\
	150 & 1 \\
	1 & 1  \\
	1 & 1  \\
	\end{bmatrix}
}.
\]
Upon introducing the twelve samples from location AT and the two trace samples into the model and observing convergence, we obtain:
\setcounter{MaxMatrixCols}{20}
\[
\scalemath{0.70}{
	\mathbf{H}_{converged}=\begin{bmatrix}
		2309.54 & 226.03 & 16.30 & 78.51 & 33.32 & 20.73 & 70.02 & 41.04 & 7.28 & 228.62 & 43.15 & 916.10 & 1.83 & 120.59 \\
		30.29 & 14.21 & 39.75 & 23.41 & 2.06 & 173.14 & 0.29 & 14.17 & 5.12 & 9.18 & 81.25 & 337.00 & 9.61 & 10.63 \\
	\end{bmatrix},		
	}
\]
\[
\scalemath{0.70}{
	\mathbf{A}_{converged} = \begin{bmatrix}
		\boldsymbol{\mathbf{\alpha}}_{e_1,1} \\
		\boldsymbol{\mathbf{\alpha}}_{e_2,1} \\
	\end{bmatrix}
	= \begin{bmatrix} 
		907.69 &  92.31 \\
		258.32 & 741.68 \\
	\end{bmatrix}
}.
\]
Note that, by design of the algorithm, the rows corresponding to $\boldsymbol{\mathbf{\alpha}}_{AT,s}$ are not updated and are not represented above. The resulting marginal posterior distributions of $\boldsymbol{\mathbf{\theta}}_{e_1,1}$ and $\boldsymbol{\mathbf{\theta}}_{e_2,1}$ are displayed in figure~\ref{ATLQ_2traces}.

\begin{figure}[H]
	\centering
	\includegraphics[scale=0.30]{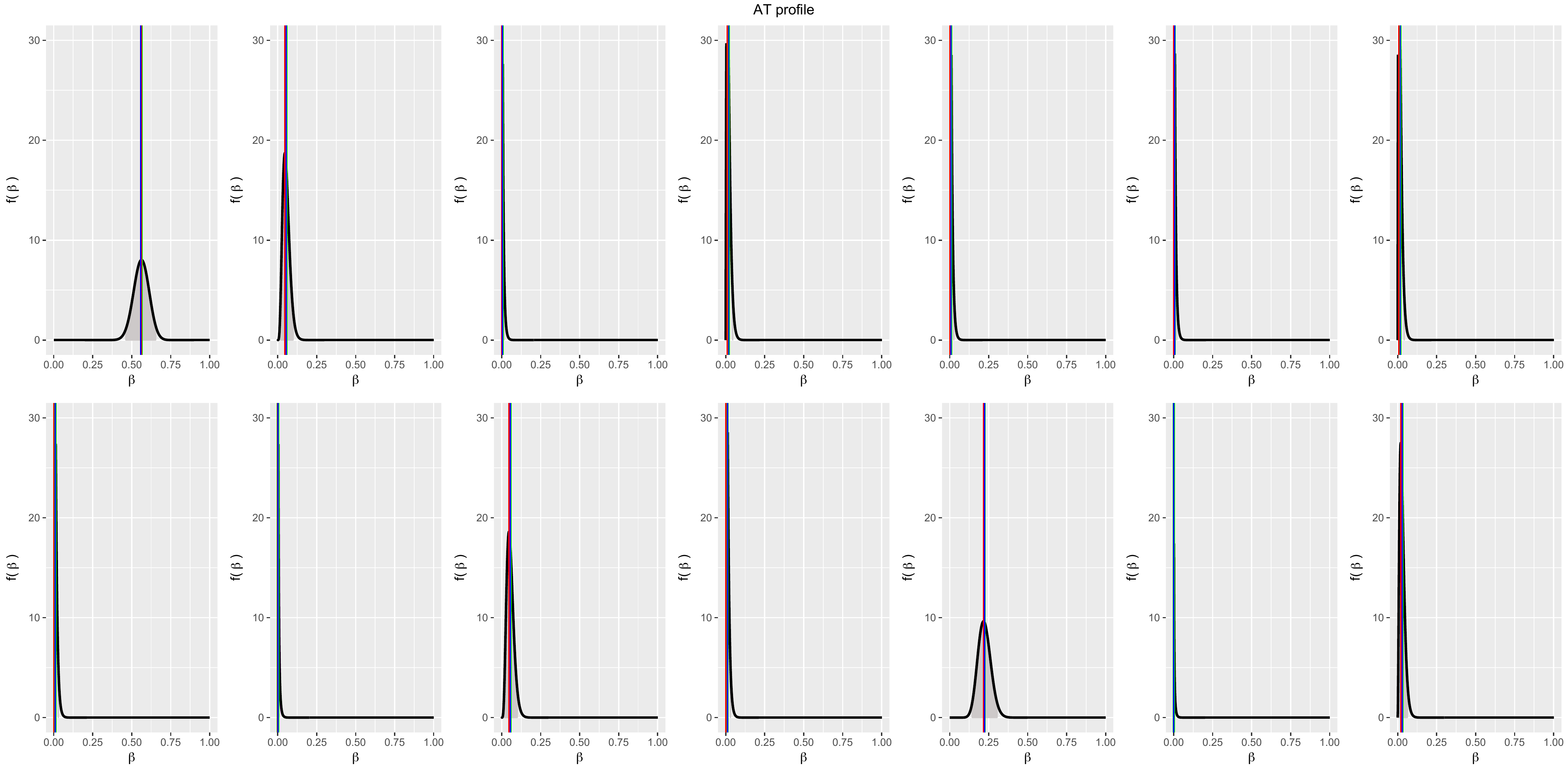} \\
	\vspace{5mm}
	\includegraphics[scale=0.30]{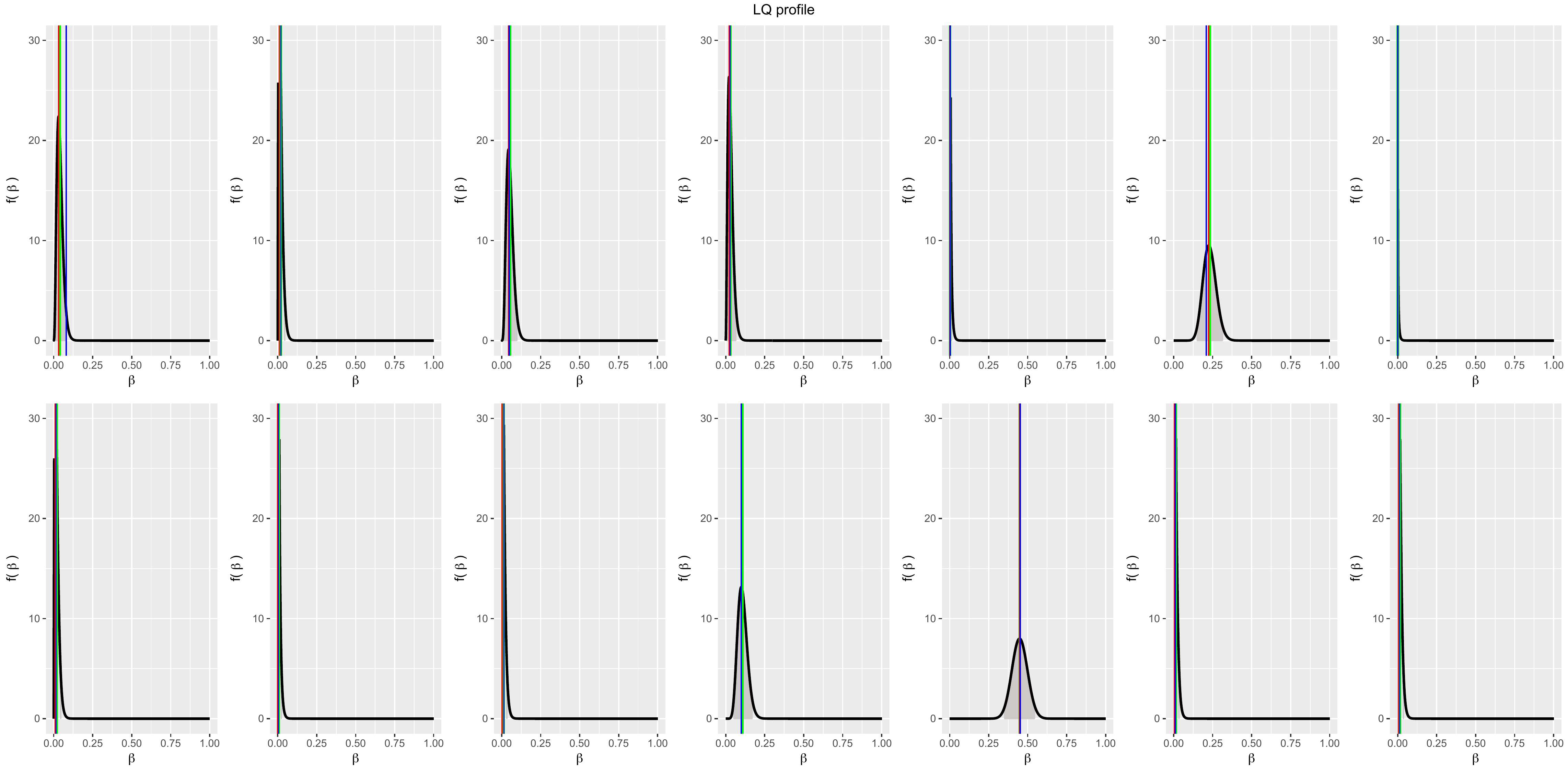}
	\caption{Plots of posterior distributions corresponding to the elements of $\mathbf{B}$ for location AT (top two rows) and location LQ (bottom two rows) when location AT is known and location LQ is unknown.  Each plot is associated with one of the fourteen particle types. The vertical blue line corresponds to the point estimates of the mixing proportions, the vertical green line corresponds to the mean of the resulting posterior distribution, and the vertical red line corresponds to the mode of the resulting posterior distribution.The grey shaded region corresponds to the 95$\%$ HPDI.}
	\label{ATLQ_profiles}
\end{figure}

Even though the algorithm is only provided with ``pure'' samples from one single source, figure~\ref{ATLQ_profiles} shows that the model remains capable of effectively extracting the profiles of both sources. That said, by comparing figures~\ref{ATLQ_both_profiles} and~\ref{ATLQ_profiles}, we note that the modes/means of the posterior distributions for the profile of source LQ are not as well aligned with the proportion estimates of the particle types  when no ``pure'' sample from source LQ is observed. We also observe that, in general, there is more uncertainty on the particle profiles in figure~\ref{ATLQ_profiles}.

Figure~\ref{ATLQ_2traces} shows that the model is able to extract accurately the mixing proportions of the locations within the dust mixture dominated by location AT, and less accurately the mixing proportions in the dust mixtures dominated by location LQ. These results seem to indicate that the lower precision (compared to the previous scenario) of the predicted particle profiles for both sources impacts the ability of the model to accurately resolve mixing proportions, in particular, in the case when the unobserved source dominates the mixture.

\begin{figure}[H]
	\centering
	\includegraphics[scale=0.25]{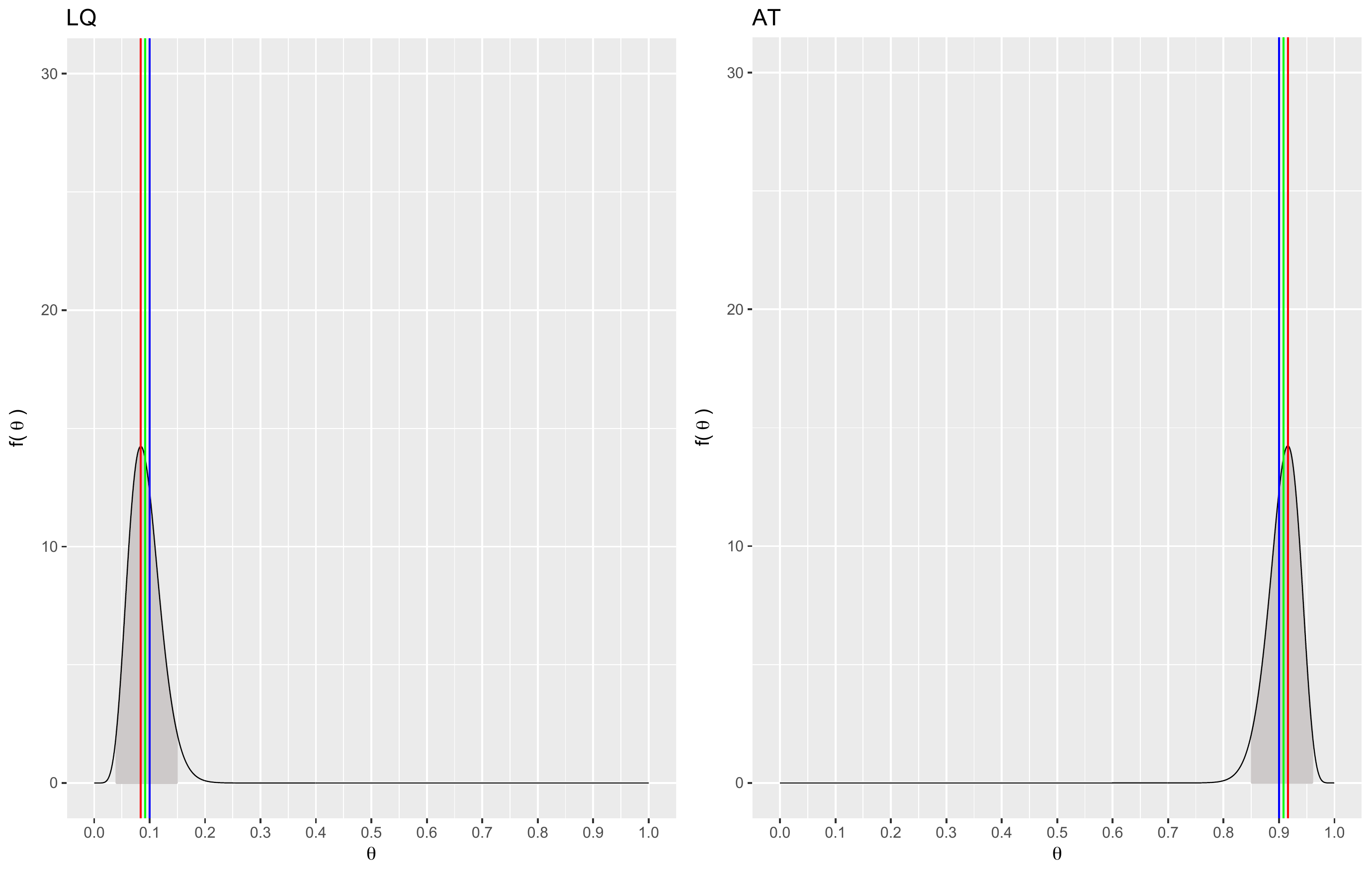} 
	\hspace{2mm}
	\includegraphics[scale=0.25]{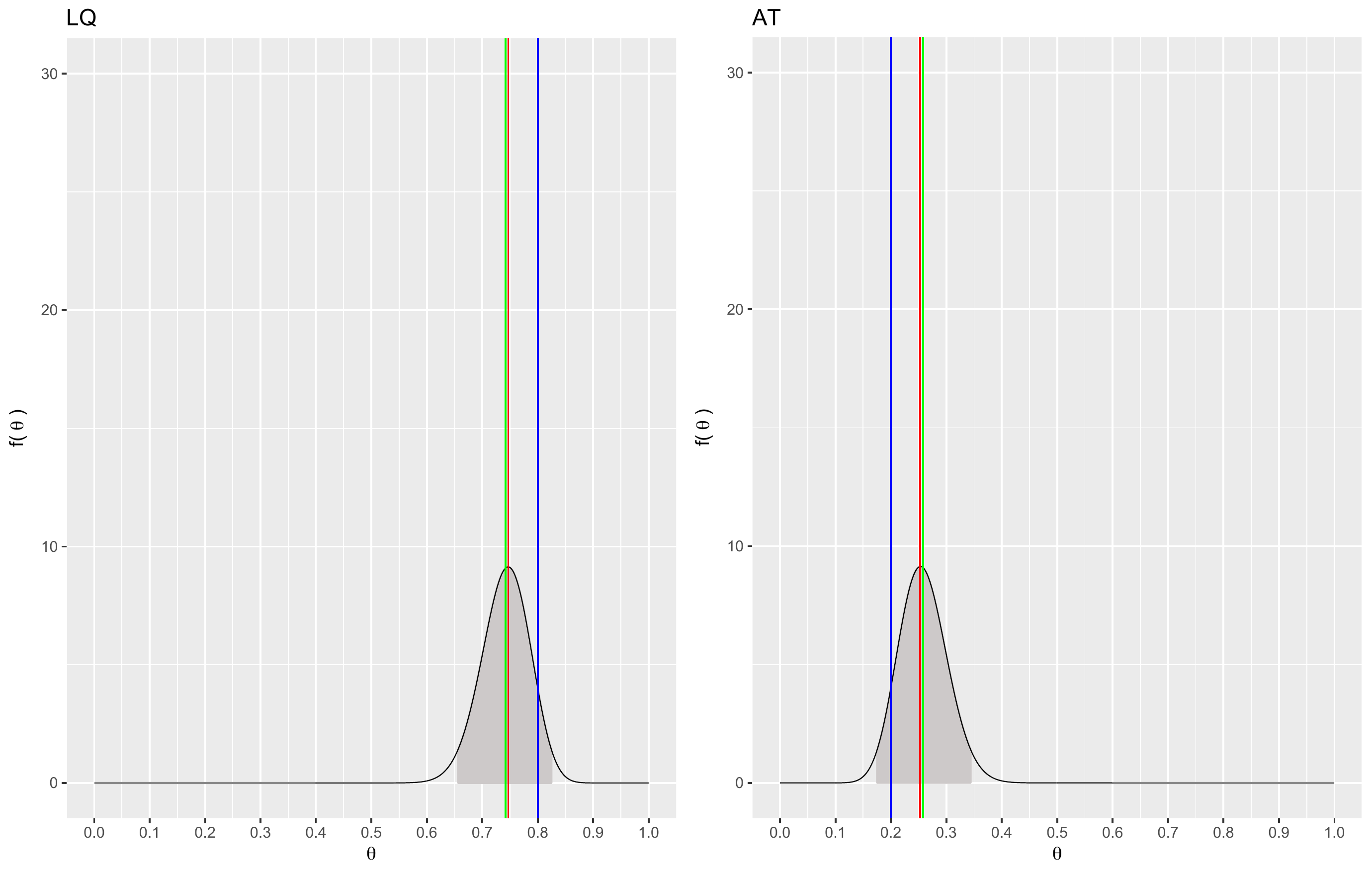} 
	\caption{Plots of posterior distributions corresponding to the elements of $\mathbf{A}$ when AT is known and LQ is unknown. The first pair of plots on the left corresponds to a trace mixture where 90$\%$ of the particles originate from (known) location AT, and 10$\%$ originate from location LQ. The second pair of plots on the right corresponds to a trace mixture where 20$\%$ of the particles originate from (known) location AT, and 80$\%$ originate from location LQ. The vertical blue line corresponds to the true mixing proportion, the vertical green line corresponds to the mean of the resulting posterior distribution, and the vertical red line corresponds to the mode of the resulting posterior distribution.The grey shaded region corresponds to the 95$\%$ HPDI.}	
	\label{ATLQ_2traces}
\end{figure}

\subsection{Deconvolving the mixtures in table~\ref{mixsamps} when only location LQ is known}
\label{LQKnown}
In the final example, we assess the model's ability to deconvolve the trace mixtures in table~\ref{mixsamps} when location LQ is known, and location AT is unknown.

Matrix $\mathbf{H}$ remains the same as in the two previous examples. The known samples that are introduced into the model are now from location LQ. We account for this difference in information by weighting the elements of matrix $\mathbf{A}$ corresponding to location LQ, rather than to location AT:
\[ 
\scalemath{0.80}{
	\mathbf{A}_{initial} = \begin{bmatrix}
		\boldsymbol{\mathbf{\alpha}}_{LQ,1} \\
		\vdots \\
		\boldsymbol{\mathbf{\alpha}}_{LQ,12} \\
		\boldsymbol{\mathbf{\alpha}}_{e_1,1} \\
		\boldsymbol{\mathbf{\alpha}}_{e_2,1} \\
	\end{bmatrix}
	= \begin{bmatrix} 
	1 & 150 \\
	\vdots & \vdots \\
	1 & 150 \\
	1 & 1  \\
	1 & 1  \\
	\end{bmatrix}
}.
\]
Upon observing convergence, we obtain the updated matrices $\mathbf{H}_{converged}$ and $\mathbf{A}_{converged}$:
\setcounter{MaxMatrixCols}{20}
\[
\scalemath{0.70}{
	\mathbf{H}_{converged}=\begin{bmatrix}
		554.24 & 55.60 & 4.59 & 19.98 & 8.84 & 7.55 & 16.54 & 12.24 & 1.98 & 57.24 & 11.96 & 210.44 & 0.83 & 31.14 \\
		314.10 & 80.56 & 195.28 & 114.51 & 12.57 & 852.21 & 2.52 & 62.72 & 20.17 & 57.10 & 401.61 & 1842.67 & 44.04 & 52.24 \\
	\end{bmatrix},		
	}
\]
\[
\scalemath{0.75}{
	\mathbf{A}_{converged} = \begin{bmatrix}
		\boldsymbol{\mathbf{\alpha}}_{e_1,1} \\
		\boldsymbol{\mathbf{\alpha}}_{e_2,1} \\
	\end{bmatrix}
	= \begin{bmatrix} 
		8934.71 & 1065.29 \\
		1821.83 & 8186.12 \\
	\end{bmatrix} \nonumber
}.
\]

\begin{figure}[H]
	\centering
	\includegraphics[scale=0.30]{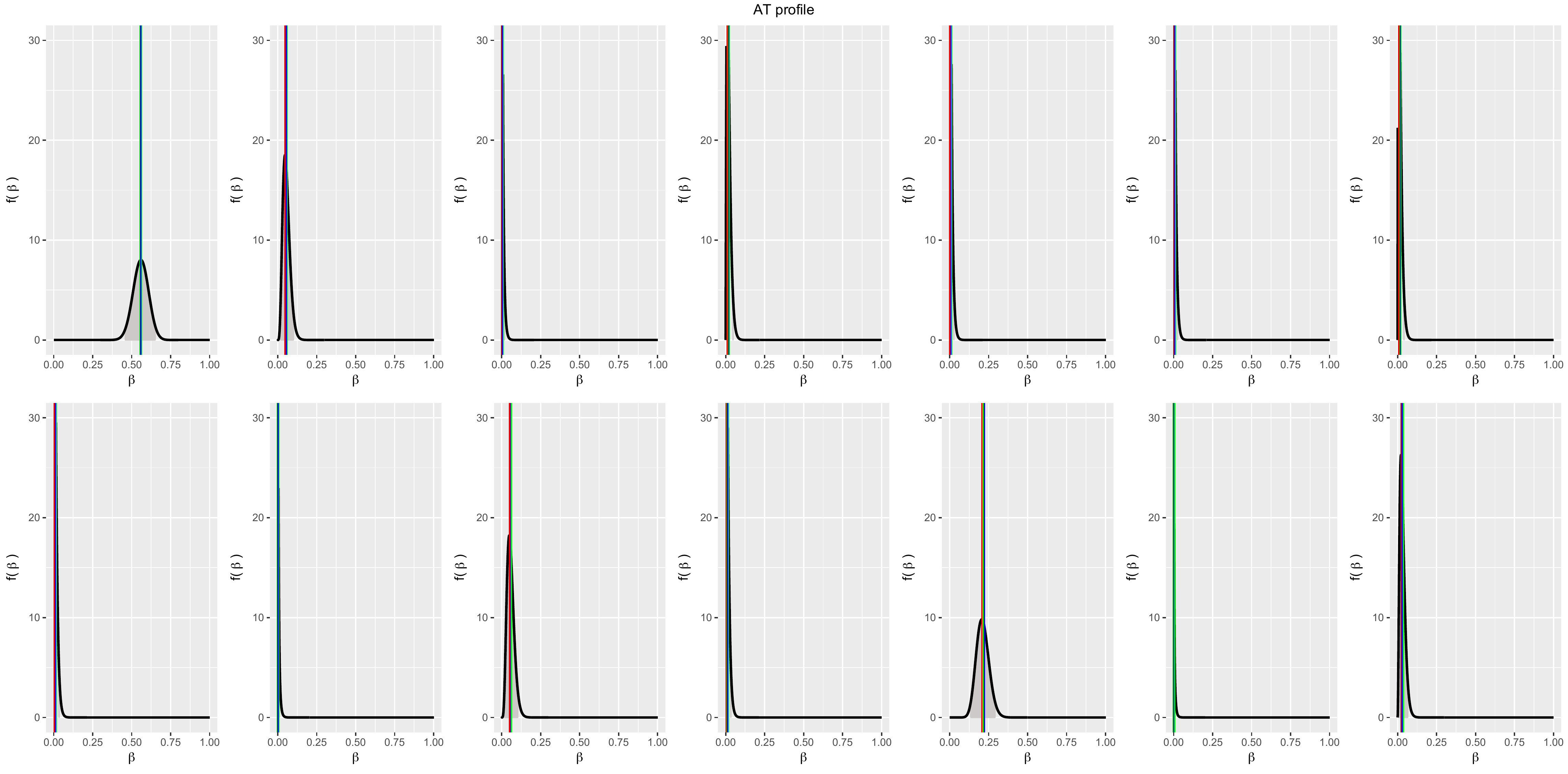}\\
	\vspace{1mm}
	\includegraphics[scale=0.30]{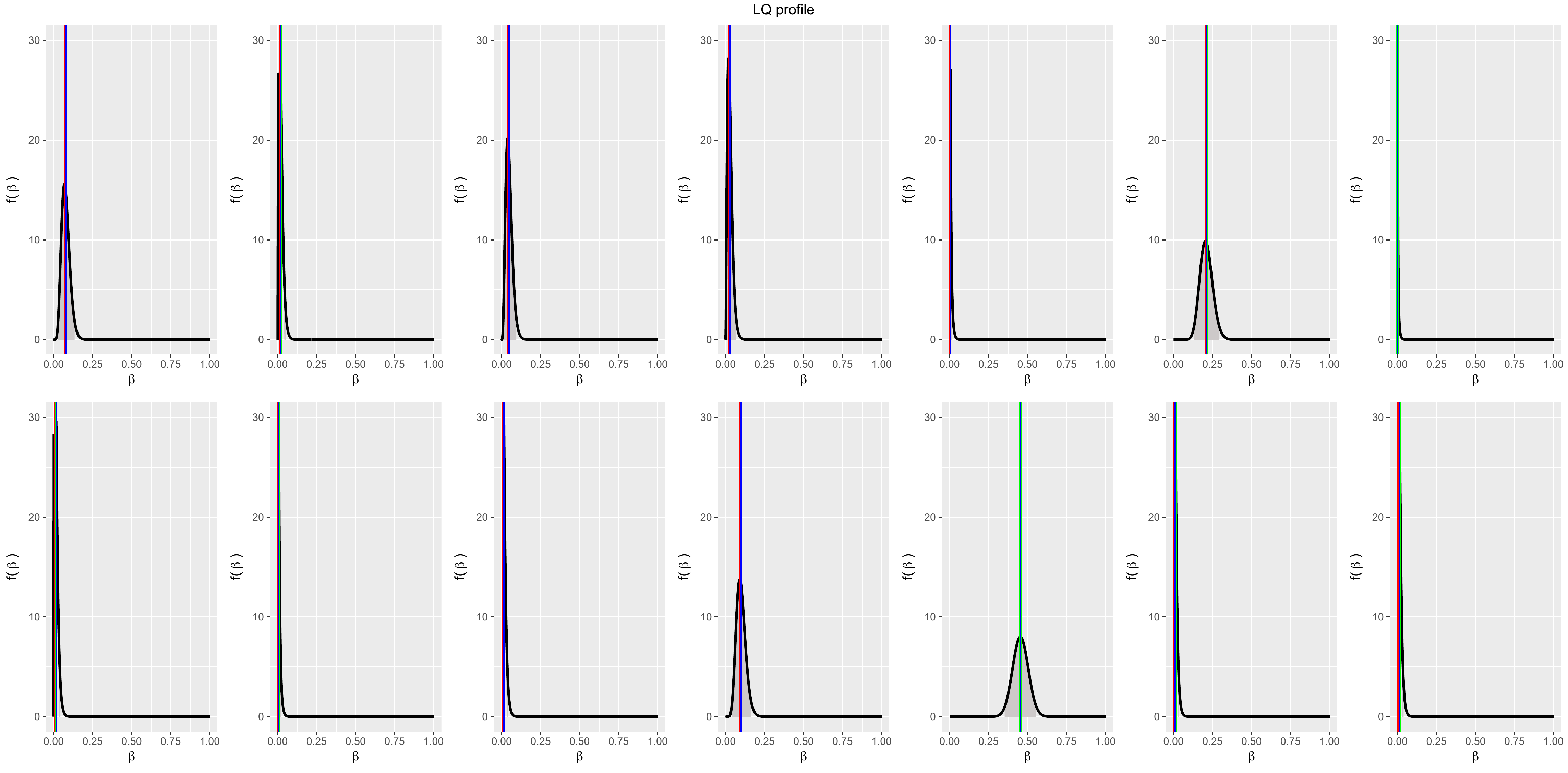} \\
	\caption{Plots of posterior distributions corresponding to the elements of $\mathbf{B}$ for location AT (top two rows) and location LQ (bottom two rows) when location AT is unknown and location LQ is known.  Each plot is associated with one of the fourteen particle types. The vertical blue line corresponds to the proportion estimates of the particle types, the vertical green line corresponds to the mean of the resulting posterior distribution, and the vertical red line corresponds to the mode of the resulting posterior distribution.The grey shaded region corresponds to the 95$\%$ HPDI.}
	\label{LQAT_Profiles}
\end{figure}

The resulting marginal posterior distributions of $\boldsymbol{\mathbf{\beta}}_{AT}$ and $\boldsymbol{\mathbf{\beta}}_{LQ}$ are displayed in figure~\ref{LQAT_Profiles}. The resulting marginal posterior distributions of $\boldsymbol{\mathbf{\theta}}_{e_1,1}$ and $\boldsymbol{\mathbf{\theta}}_{e_2,1}$ are displayed in figure~\ref{LQAT_2traces}.

As in the two previous examples, figure~\ref{LQAT_Profiles} shows that the model is able to extract the two location profiles present in the evidentiary samples: the mean and mode of the distributions prove to be reasonably similar to the proportion estimates of particle types. However, when contrasting figures~\ref{ATLQ_profiles} and ~\ref{LQAT_Profiles}, we see that the precision of the determination of the particle profiles in figure~\ref{LQAT_Profiles} is greater than in the previous example when AT was known and LQ unknown.

Contrary to the previous example, in the situation where only location LQ is known, the model is able to deconvolve both trace mixtures appropriately. It is not clear why the model behaves differently in these two different examples. We suspect that, since there are more overall particles from location AT in the mixtures (90\% of AT in the first mixture and 20\% of AT in the second mixture vs. 10\% and 80\% for LQ), the model may be able to learn the particle profile of location AT with greater accuracy even though ``pure'' samples from location AT are not provided. 

\begin{figure}[H]
	\centering
	\includegraphics[scale=0.25]{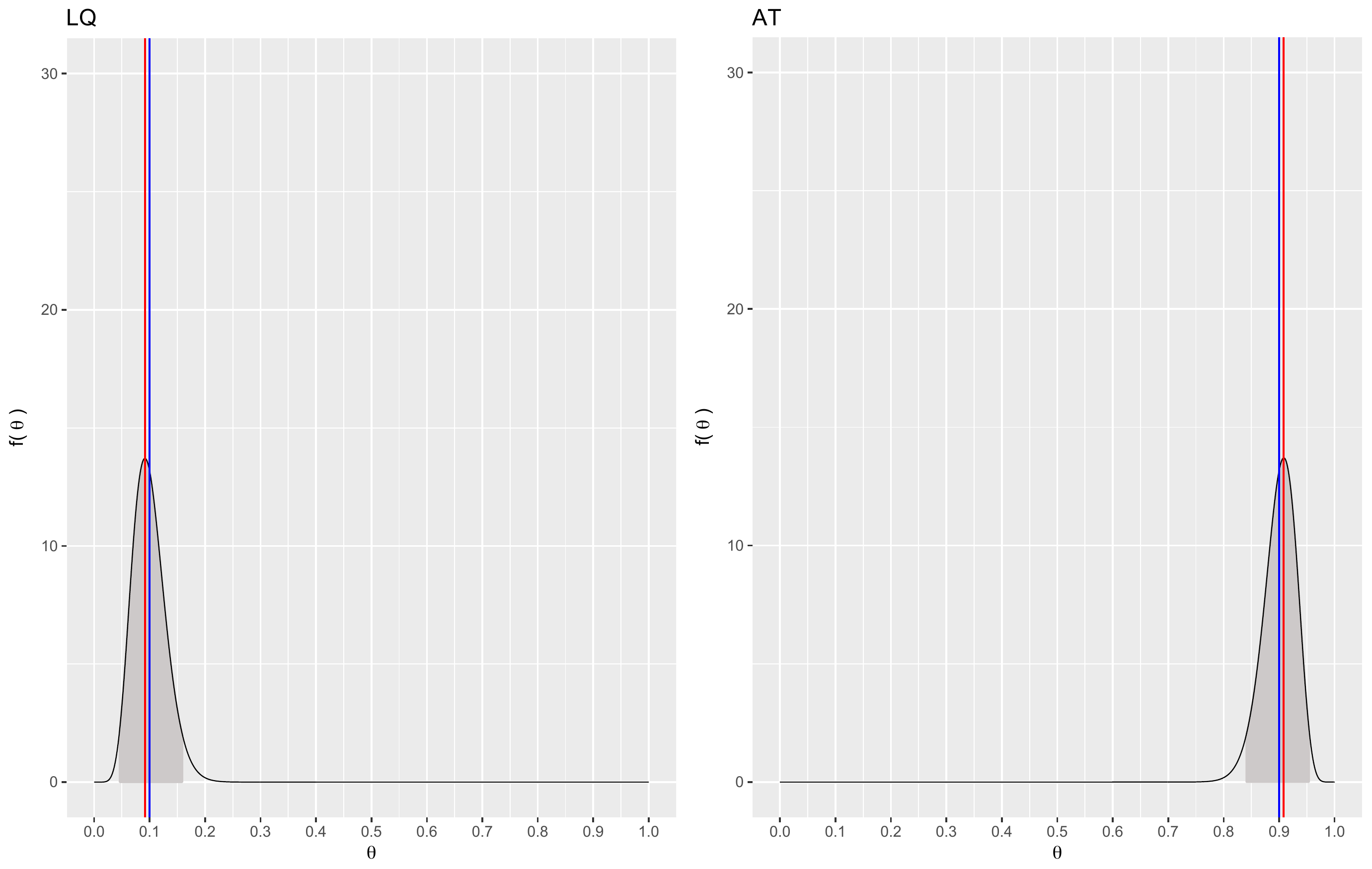} \hspace{2mm}
	\includegraphics[scale=0.25]{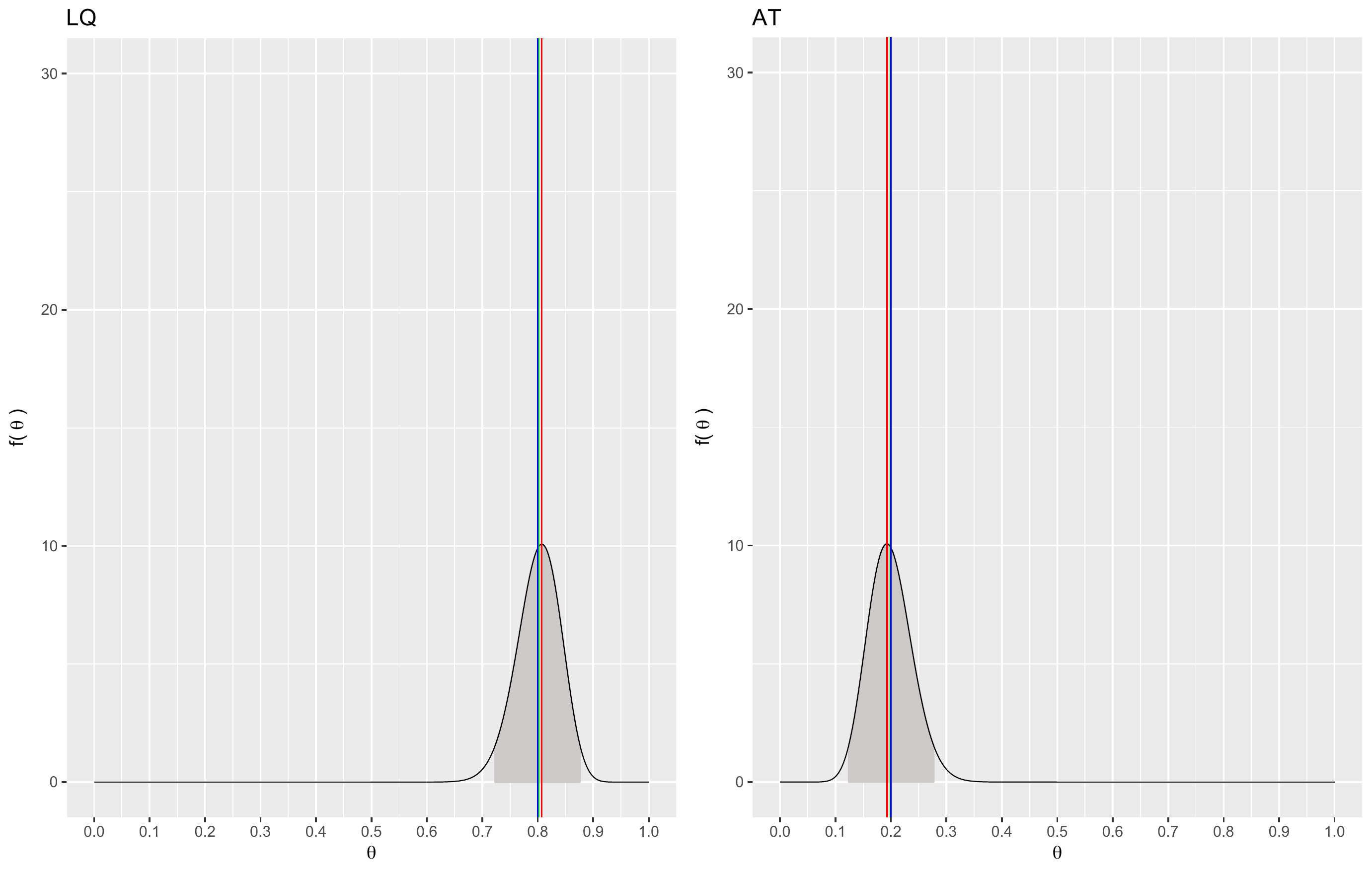} 
	\caption{Plots of the posterior distributions corresponding to the elements of $\mathbf{A}$ when location LQ is known and location AT is unknown. The first pair of plots  on the left corresponds to a mixture where 90$\%$ of the particles originate from location AT, and 10$\%$ originate from (known) location LQ. The second pair of plots on the right corresponds to a mixture where 20$\%$ of the particles originate from known location AT, and 80$\%$ originate from (known) location LQ. The vertical blue line corresponds to the known mixing proportion, the vertical green line corresponds to the mean of the resulting posterior distribution, and the vertical red line corresponds to the mode of the resulting posterior distribution.The grey shaded region corresponds to the 95$\%$ HPDI.}
	\label{LQAT_2traces}
\end{figure} 

\section{Model performance}
\label{sec:sims}
The results presented in sections~\ref{ATKnown} and~\ref{LQKnown} show that the performance of the model may differ depending on the input data.
A search of the literature for discussions on the identifiability of LDA models indicated that this issue has only been considered by very few authors \citep{Rabani:2013, Vandermeulen:2015} without providing satisfactory solutions. Furthermore, we are not aware of published LDA-based models where the models' outputs are compared to ground truth. While many authors propose different flavours of LDA-based models, we have yet to find a publication whose authors test their model on a simulated corpus (whose proportions of topics and topic profiles are known) to get an idea of the general accuracy and identifiability of their models.

To study the performance of our algorithm, we simulate dust mixtures in which the particle counts and mixing proportions vary. Recall that by assumption (f), in section~\ref{model_assumptions}, we have, at most, one unknown location in a sample obtained from any given evidentiary object and that this assumption is compatible with an operational use of the model in the forensic context. We choose to simulate situations where we observe dust mixtures composed of two sources, AT and LQ. In all situations, source AT is known and can be sampled from, while source LQ is not known, and its profile is left to be learned by the model from the dust mixtures. These situations correspond to the example in section~\ref{ATKnown}, which was the one where our model had the most difficulties. We consider two situations:
\begin{enumerate}[(a)]
	\item a single set of trace objects representing a single mixture of AT and LQ, originating from a single sampling location, is observed.
	\item two sets of trace objects, representing two different mixtures of AT and LQ, originating from two different sampling locations, are observed.
\end{enumerate}
We do not present the results of the situation in which ``pure'' dust samples can be obtained for both sources in the mixtures. The performance of the model in this situation is analogous to that presented in section~\ref{bothknown_experiment}, and is, overall, uninteresting.

We use point estimates for the proportions of the particle types for sources AT and LQ (obtained using the data presented in tables~\ref{ATsamps} and~\ref{LQsamps}) to generate dust samples from these sources before mixing them. In each simulation, we consider trace sets that consist in five samples of mixtures of dust from sources AT and LQ, and a control set consisting in twelve samples obtained from location AT. The number of samples was selected to correspond to a realistic forensic scenario where both trace and control locations would be sampled several times to study their respective variability, but where the number of samples would not be unrealistic, or so high that their examination would be too time consuming. The mixing proportions of the two sources in situation (a) varies between simulations, such that $\boldsymbol{\mathbf{\theta}}_{e_1,s}$ takes the values in the following column matrix:
\[
\scalemath{0.80}{
	\boldsymbol{\mathbf{\theta}}_{e_1,s} \in \begin{bmatrix}
	0.10 & 0.20 & 0.25 & 0.33 & 0.50 & 0.67 & 0.75 & 0.80 & 0.90\\
	0.90 & 0.80 & 0.75 & 0.67 & 0.50 & 0.33 & 0.25 & 0.20 & 0.10 \end{bmatrix}
}, \ \ \text{for} \ s \in \{1,2,3,4,5\}.
\]
 The mixing proportions associated with situation (b) remain the same as in situation (a) for the first sampling location and have been set to $\boldsymbol{\mathbf{\theta}}_{e_2,s}= (0.51,0.49)$, for $s \in \{1,2,3,4,5\}$, for the second sampling location. 
 
 For each of the nine mixing proportions considered for $\boldsymbol{\mathbf{\theta}}_{e_1,s}$, the particle count in each dust mixture varies so that $N_s \in \{100, 300, 500, \dots,$ $1500, 1700, 1900\}$, $s \in \{1,2,3,4,5\}$. This results in a total of 90 different simulations for each situation. Each simulation is repeated ten times, and the average model performance is evaluated. 
 

\subsection{Simulation results}
\subsubsection{One set of trace objects}

The results obtained from the simulations conducted for situation (a) are presented in figures~\ref{PredictedProfiles_OneTrace_ATKnown_AT}, \ref{PredictedProfiles_OneTrace_ATKnown_LQ}, and \ref{PredictedProportions_OneTrace_ATKnown}. Figures~\ref{PredictedProfiles_OneTrace_ATKnown_AT} and \ref{PredictedProfiles_OneTrace_ATKnown_LQ} show the average predicted values of $\beta_{m, t}$, $m\in \{AT, LQ\}$, $t \in \{1, \hdots, 14\}$ as a function of the number of particles present in each sample, and as a function of the true mixing proportion of source LQ present in the trace samples. From figure~\ref{PredictedProfiles_OneTrace_ATKnown_AT}, we see that the model is able to accurately extract the profile of known source AT, which is not surprising given that ``pure'' samples of location AT were provided. However, given a single trace sampling location, we see from figure~\ref{PredictedProfiles_OneTrace_ATKnown_LQ} that the model struggles to extract the profile of the unknown source, LQ. 

\begin{figure}[H]
	\centering
	\includegraphics[scale=0.40]{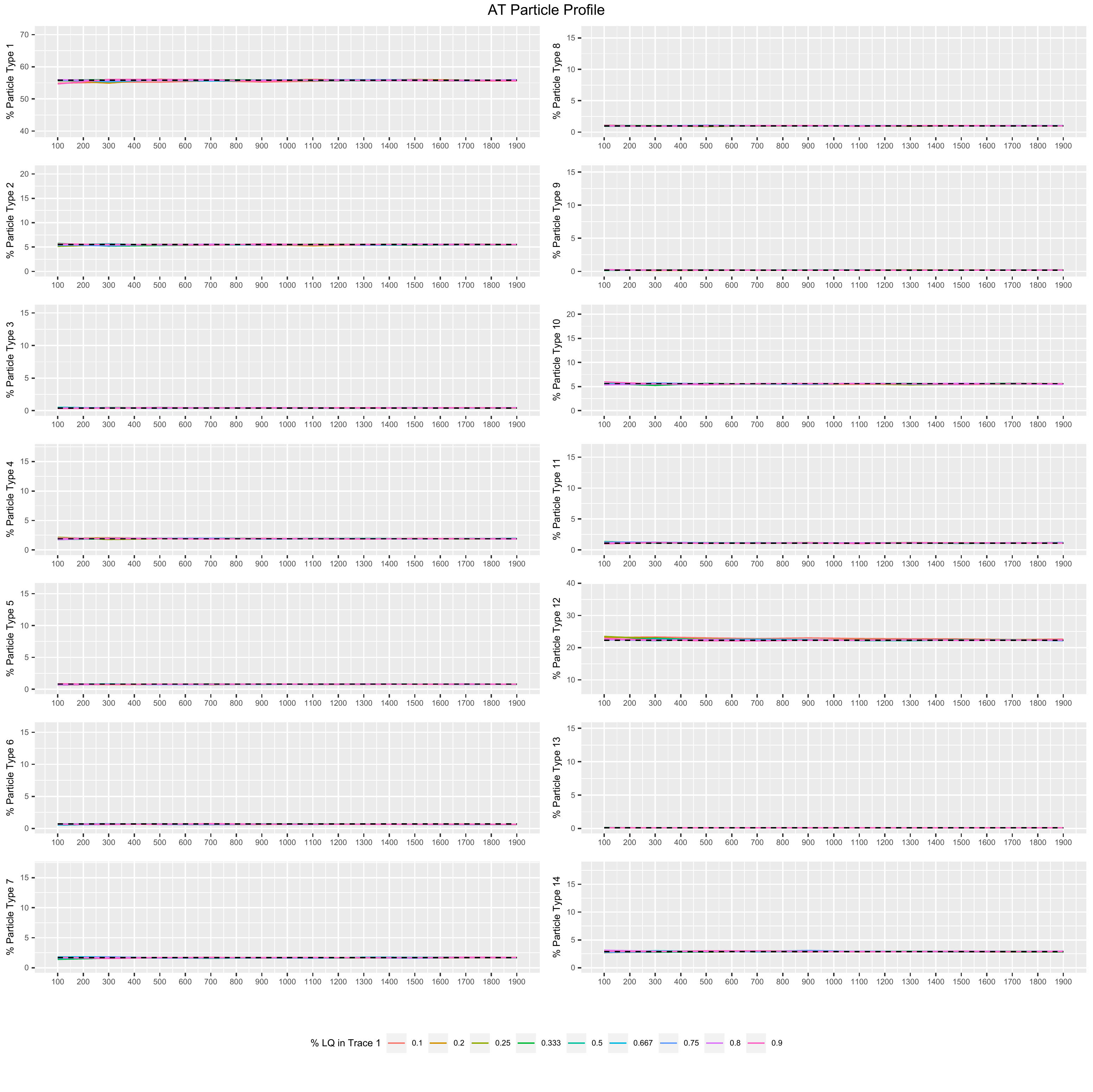}
	\caption{Predicted profile of source AT when source AT is known and able to be sampled from, and when a single set of trace objects is observed from a single sampling location. Predicted particle proportions are obtained using the average value of the means of the posterior distributions of each $\beta_{AT, t}$ obtained from ten replications of the simulations.  \label{PredictedProfiles_OneTrace_ATKnown_AT}}
\end{figure}

\begin{figure}[H]
	\centering
	\includegraphics[scale=0.40]{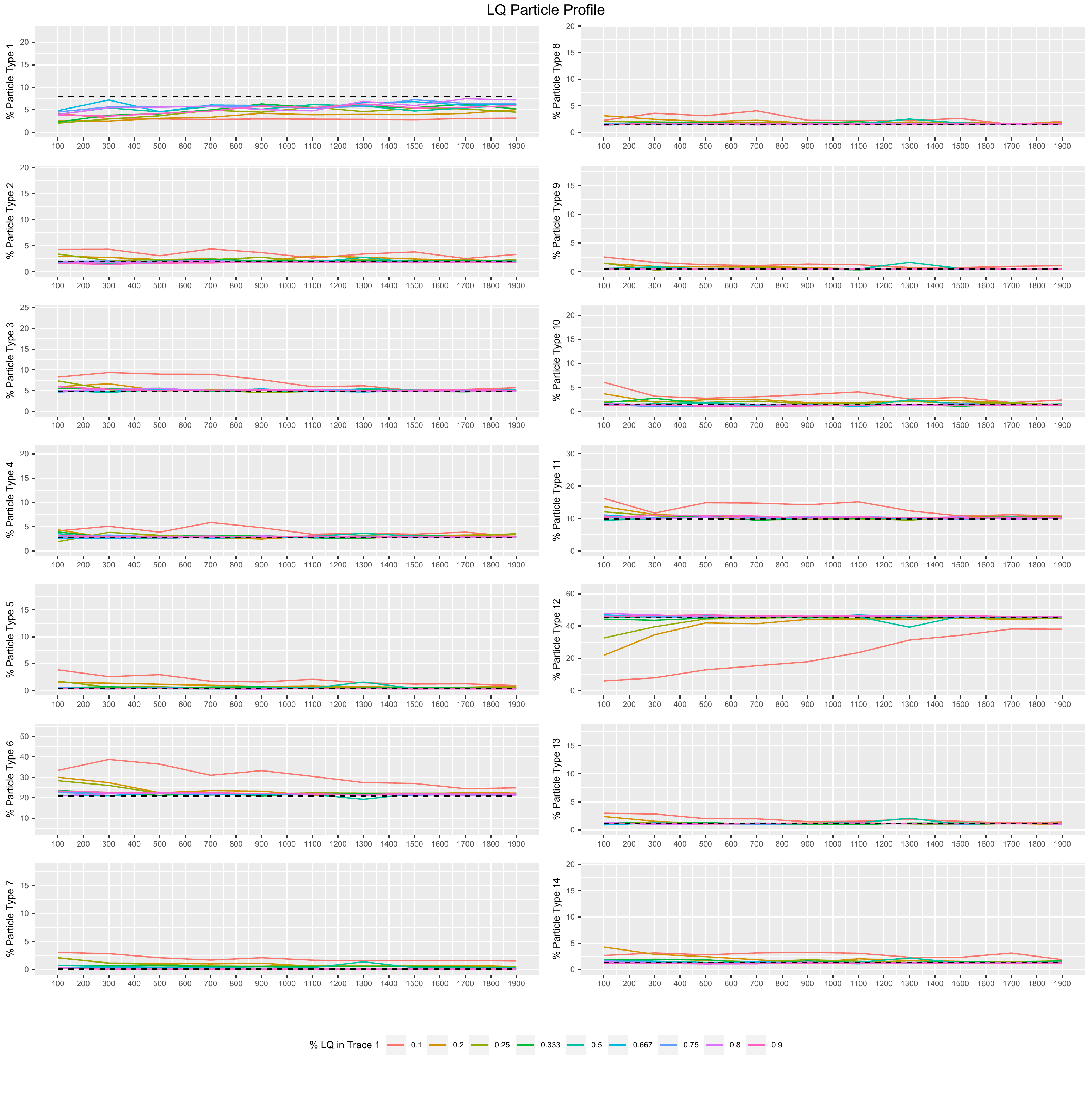}
	\caption{Predicted profile of source LQ when source AT is known and able to be sampled from, and when a single set of trace objects is observed from a single sampling location. Predicted particle proportions are obtained using the average value of the means of the posterior distributions of each $\beta_{LQ, t}$ obtained from ten replications of the simulations.  
	\label{PredictedProfiles_OneTrace_ATKnown_LQ}}
\end{figure}

Figure~\ref{PredictedProfiles_OneTrace_ATKnown_LQ} shows two general aspects of the behaviour of the model. First, the accuracy of the prediction of the location particles' profiles improves as the number of particles present in the samples increases. Second, larger proportions of the unknown source in the trace mixture results in more accurate predictions of its profile. These two aspects of the behaviour of the model are not surprising as, in both cases, the predictions improve with the amount of information regarding the profile of the unknown source. More interestingly, comparing figures~\ref{PredictedProfiles_OneTrace_ATKnown_AT} and~\ref{PredictedProfiles_OneTrace_ATKnown_LQ} shows that the model has consistent difficulties to accurately predict the proportions of the particles types that have non-zero proportions in both profiles. In particular, it appears that the prediction is worse for particle types where the ratio of proportions between AT and LQ is largely in favour of AT, such as particle types 1 (0.56 vs. 0.08), 2 (0.055 vs. 0.02), 7 (0.017 vs. 0.001) or 10 (0.056 vs. 0.014). When the ratio is largely favourable to LQ, such as in particle types 6 (0.007 vs. 0.21) or 12 (0.22 vs. 0.45), predictions improve as the amount of information available regarding LQ increases. This observation questions the general identifiability of LDA models, in particular when all topic profiles have to be learned from the data as in text mining. 

Figure~\ref{PredictedProportions_OneTrace_ATKnown}(top) shows the predicted mixing proportions of the unobserved source LQ in the trace mixture as a function of the number of particles present in each sample, and as a function of the true mixing proportion of source LQ present in the trace samples. Figure~\ref{PredictedProportions_OneTrace_ATKnown}(bottom) shows the deviation of the predicted mixing proportions from the true proportion. In this case, there does not appear to be a clear trend in the ability of the model to predict the mixing proportions as a function of the proportion of the unknown source in the trace mixture. We do note, however, that the performance of the model does improve as the number of particles present in the samples increases. Overall, it does not appear that the mean is a better predictor than the mode, and reciprocally. As mentioned in section~\ref{ATKnown}, the inability of the model to resolve the particle profiles of the unknown location is unsurprisingly related to its ability to resolve the mixing proportion. 

\begin{figure}[H]
	\centering
	\includegraphics[scale=0.28]{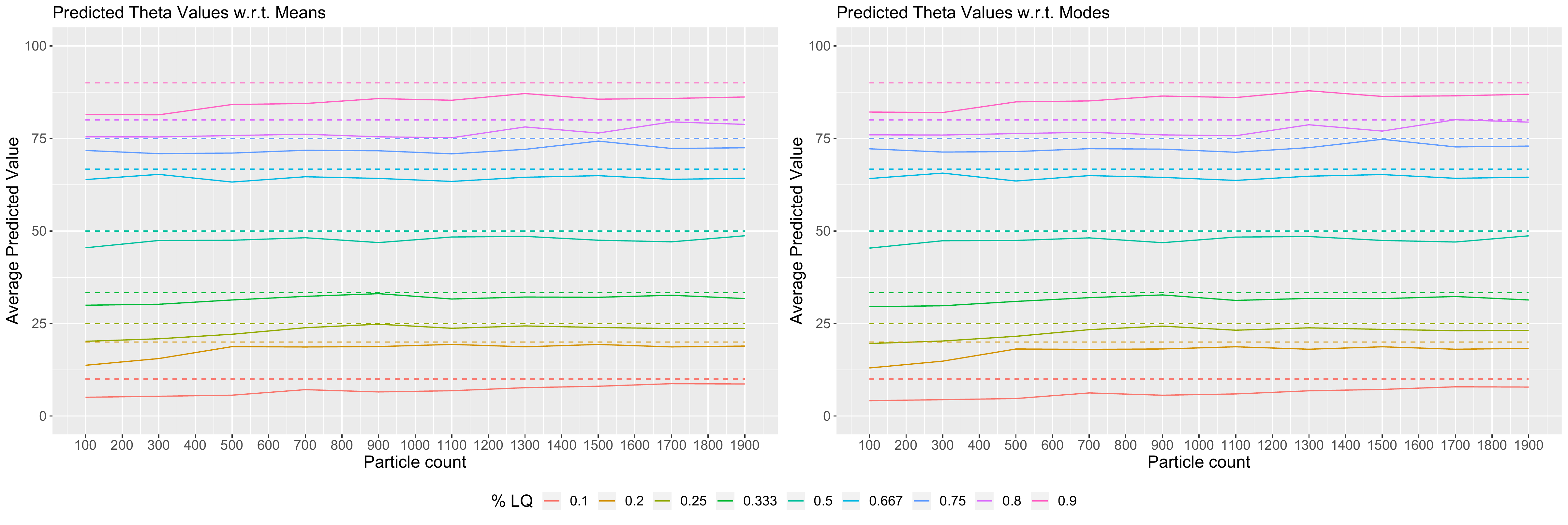}
	\includegraphics[scale=0.28]{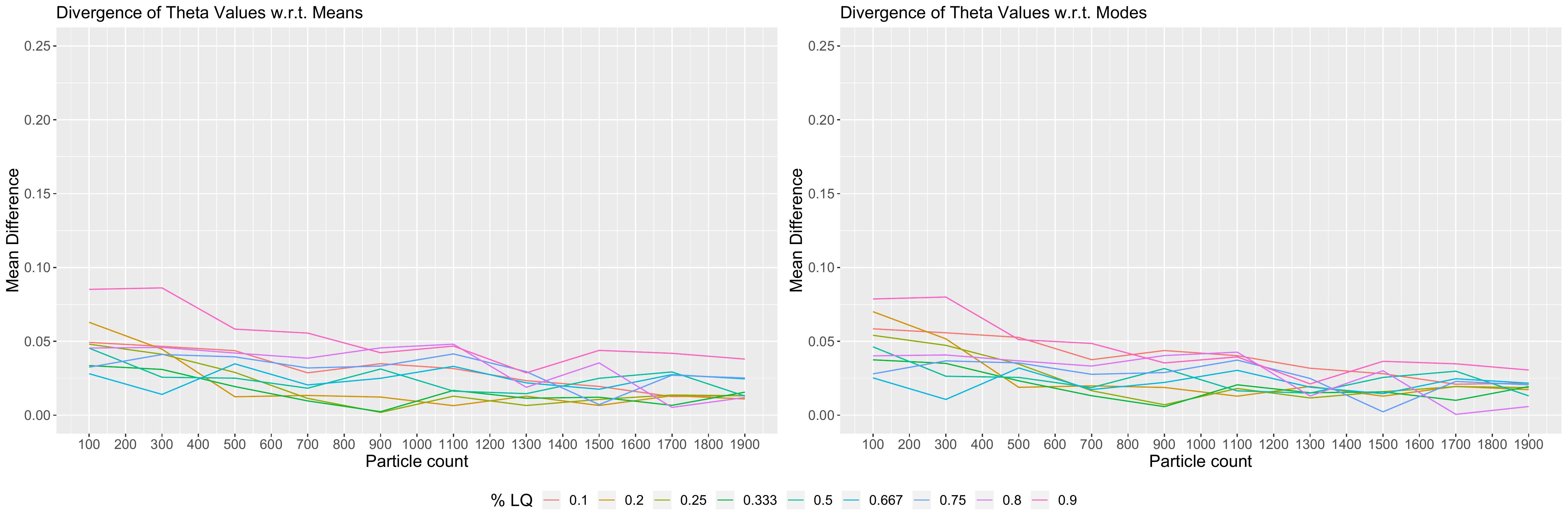}
	\caption{Predicted proportion of unobserved source LQ using the average value of the means (top left) and modes (top right) of the posterior distributions of each $\theta_{e_u, s, LQ}$, $u \in \{1,2\}$, $s \in \{1, \hdots, 5\}$ obtained from each simulation, and the divergence of the predicted values from the true mixing proportion (bottom row). \label{PredictedProportions_OneTrace_ATKnown}}
\end{figure}

\subsubsection{Two sets of trace objects}

\begin{figure}[H]
	\centering
	\includegraphics[scale=0.4]{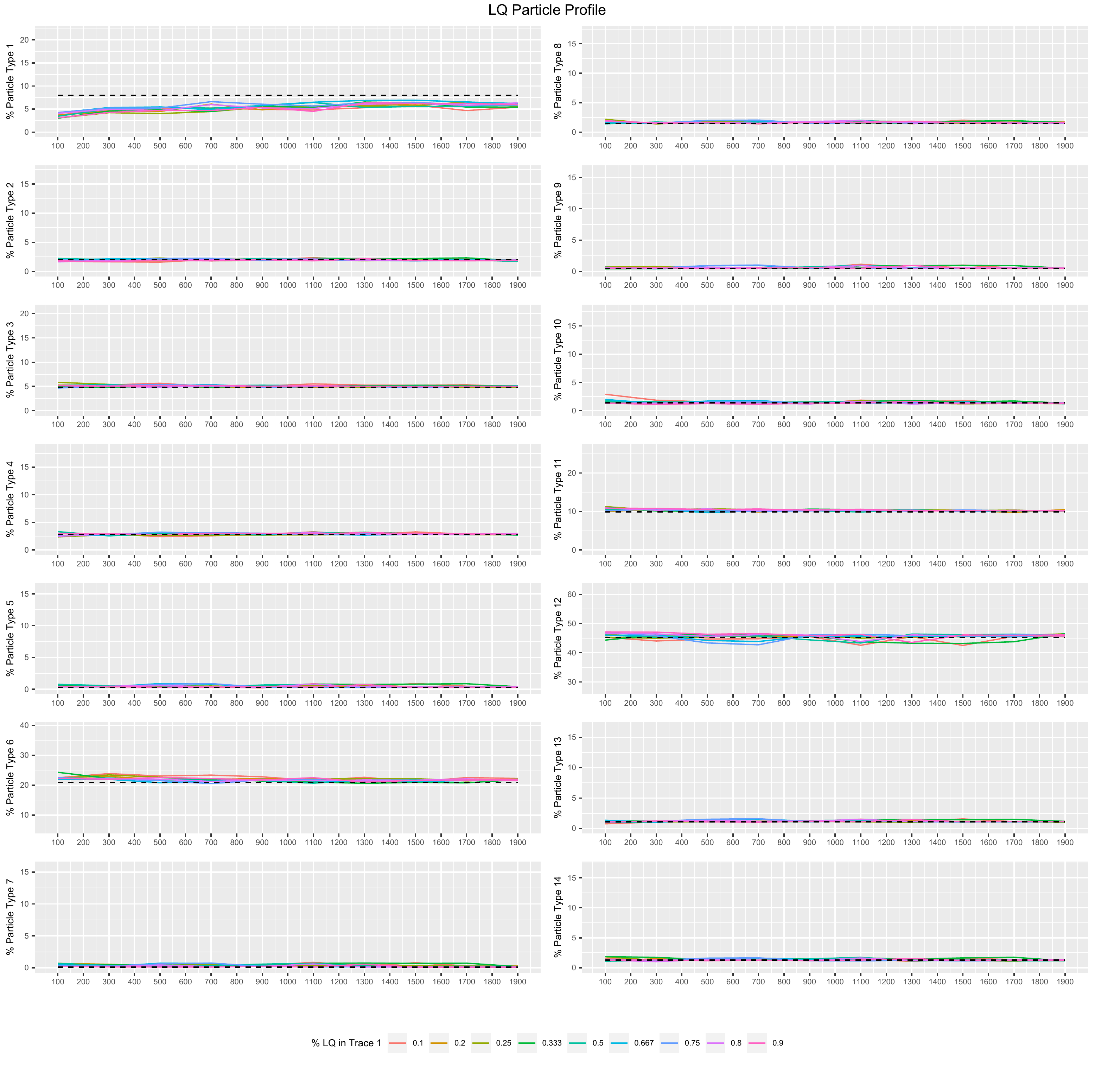}	
	\caption{Predicted profile of source LQ when source AT is known and able to be sampled from, and when two sets of trace objects are observed from two distinct sampling locations. Predicted particle proportions are obtained using the average value of the means of the posterior distributions of each $\beta_{LQ, t}$ obtained from ten replications of the simulations.    \label{PredictedProfiles_TwoTraces_ATKnown}}
\end{figure}

The results obtained from the simulations conducted for situation (b) are presented in figures~\ref{PredictedProfiles_TwoTraces_ATKnown} and \ref{PredictedProportions_TwoTraces_ATKnown}. Figure~\ref{PredictedProfiles_TwoTraces_ATKnown} shows the average predicted values of $\beta_{LQ,t}$, $t\in\{1, \hdots, 14\}$ as a function of the number of particles present in each sample, and as a function of the true mixing proportion of source LQ present in the trace samples. As previously, the model is able to accurately predict the profile of known source AT; therefore, the results are not displayed below and the reader can refer to figure~\ref{PredictedProfiles_OneTrace_ATKnown_AT}. The results in figure~\ref{PredictedProfiles_TwoTraces_ATKnown} illustrate that the presence of two sets of trace objects obtained from two distinct sampling locations greatly improves the model's ability to predict the profile of the unknown source in terms of accuracy and precision. We note that the model still has difficulties to predict particle type 1; however, there is greater precision in the prediction. 

\begin{figure}[H]
	\centering
	\includegraphics[scale=0.28]{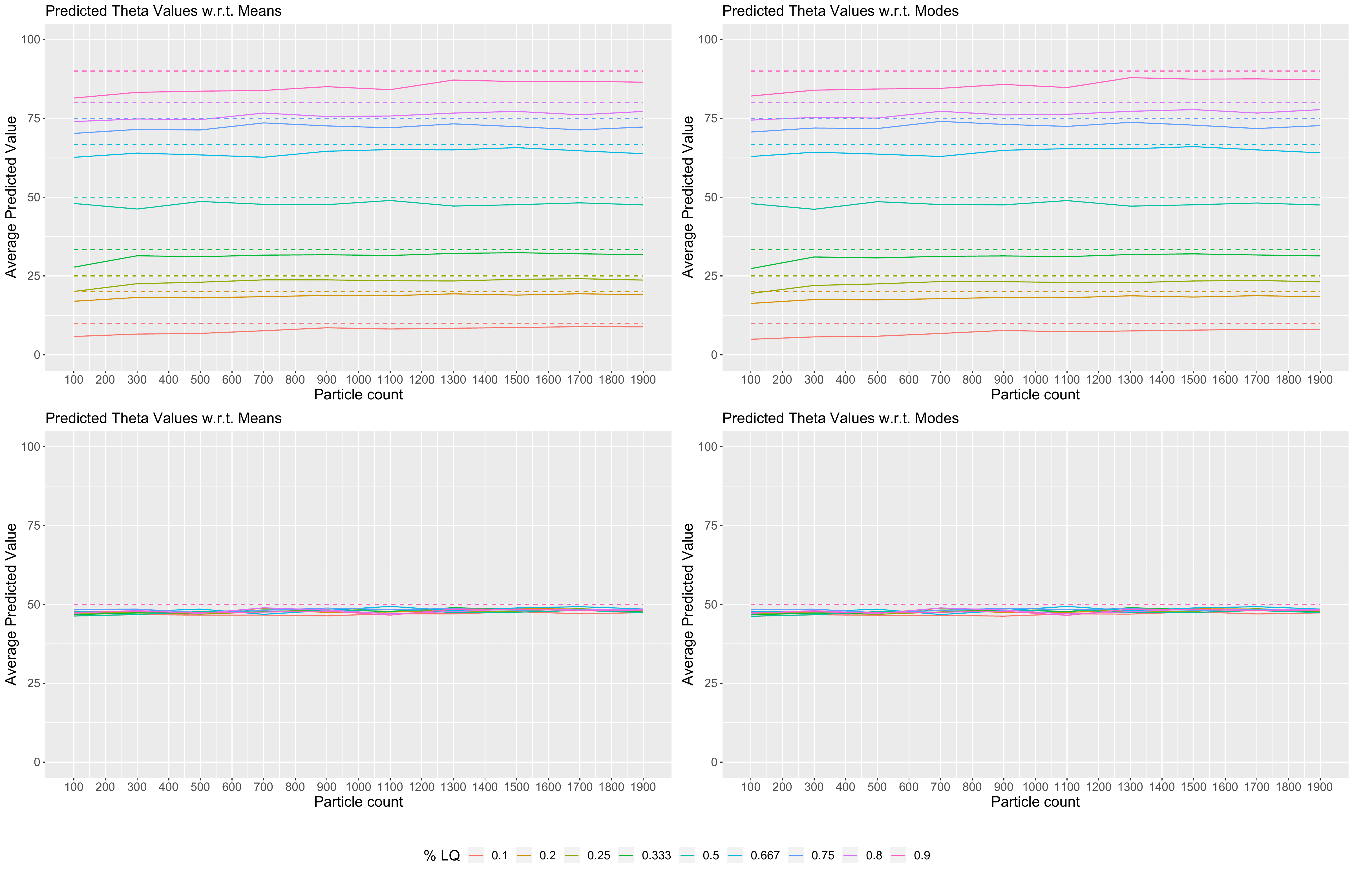} \\
	\includegraphics[scale=0.28]{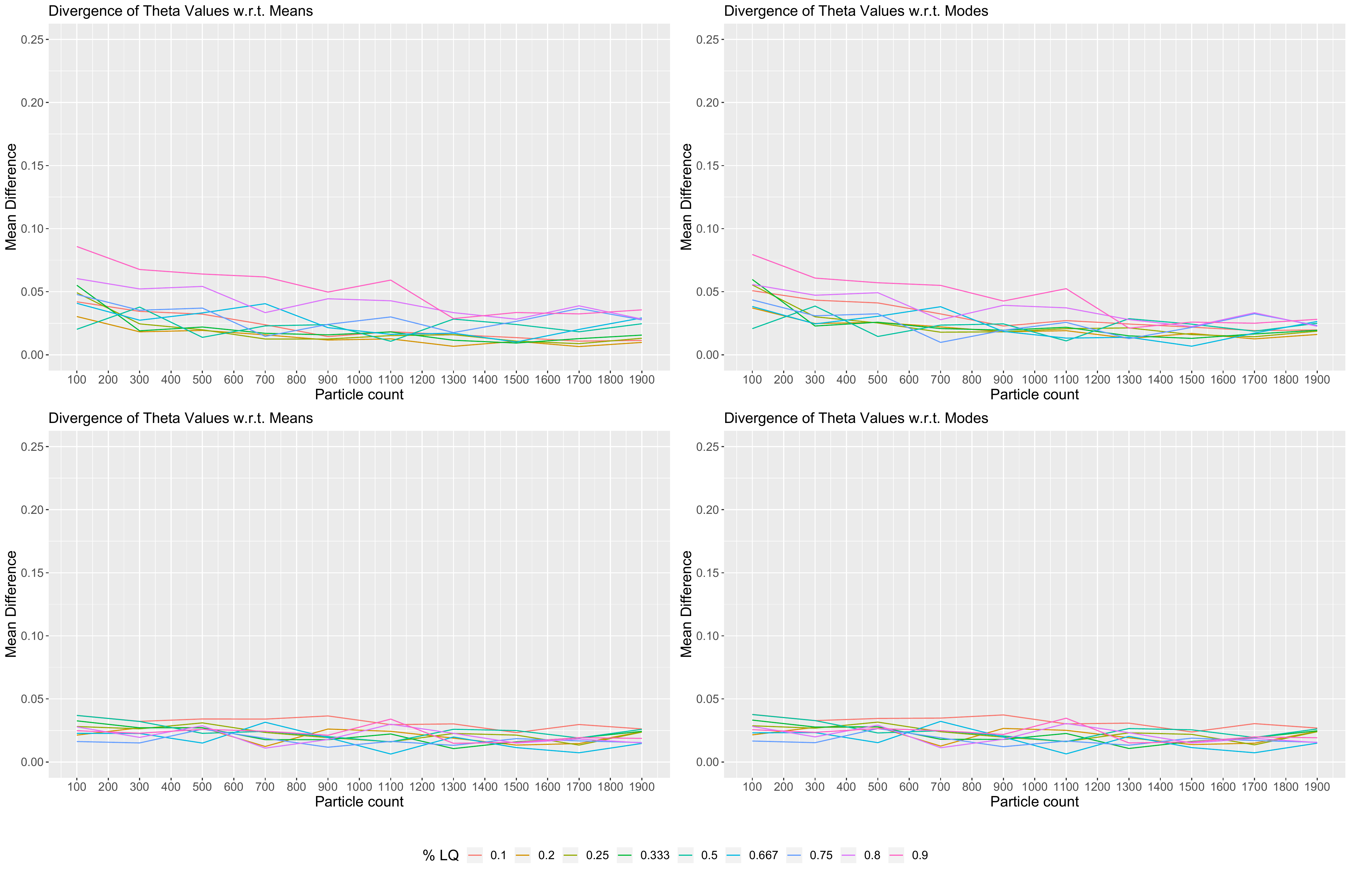}
		\caption{Predicted proportion of unobserved source LQ in each trace using the average value of the means (top two rows, first column) and modes (top two rows, second column) of the posterior distributions of each $\theta_{e_u, s, LQ}$, $u \in \{1,2\}$, $s \in \{1, \hdots, 5\}$ obtained from our  simulations, and the divergence of the predicted values from the true mixing proportion (bottom two rows). \label{PredictedProportions_TwoTraces_ATKnown}}	
\end{figure}

Figure~\ref{PredictedProportions_TwoTraces_ATKnown} shows the predicted mixing proportion of the unobserved source, LQ, in the first trace mixture (top row) and the second trace mixture (second row), as a function of the number of particles present in each sample, and as a function of the true mixing proportion of source LQ present in the first trace sample. In addition, figure~\ref{PredictedProportions_TwoTraces_ATKnown} shows the deviation of the predicted mixing proportions from the true proportions for the first set of traces (third row) and for the second set of traces (fourth row). In the situation where two sets of trace objects are observed, we see the same convergence in the ability of the model to predict the mixing proportions as in the situation with only one set of trace objects: the convergence is a function of the particle count. However, in this case, we note that the model's ability to correctly predict the proportion of LQ in the trace mixture appears to be a function of the true proportion of LQ present in the mixture: the larger the contribution of the unknown source to the dust mixture, the more the model struggles to extract the true proportion. We believe that this is due to the greater potential for error in mixtures where there is a larger proportion of the unknown source present in the mixture.

\subsection{A note on performance}
The main difference between our model and that originally proposed by \citet{Blei:2003} is that we use asymmetric Dirichlet distributions instead of symmetric ones (\citet{Blei:2003}, p. 1006, footnote 2). Overall, we are not sure whether the lack of accuracy of our model originates from a general lack of identifiability of LDA models; from the large number of parameters to be assigned in $\mathbf{H}$ and $\mathbf{A}$, which may require a much larger number of samples than the number considered by our simulations; or from some instability of the numerical optimisation methods used in the M-step of our algorithm. We have found little to no published material reporting on the performance of LDA models. Further investigations and future developments of our method may involve assigning the hyper-parameters of our model using the method of moments, or implementing a Gibbs sampler or an Approximate Bayesian Computation algorithm to obtain posterior samples of parameters of interest. Exploring the method of moments may allow to set restrictions that will help with the identifiability of the model and has the potential to fully recover the parameters of the model, given that the necessary assumptions are fulfilled \citep{Anandkumar:2012}.

Nevertheless, we want to stress that, from an operational point of view, our model performs well. In an operational situation, investigators or fact-finders will be far more concerned with information on the presence/absence of dust from a particular source in a mixture, with the ball-park contribution of this source (i.e., minimal vs. large) to the mixture and with any interesting characteristics that the profile of this source might have (e.g., a large proportion, or the complete absence, of a particular particle type) than with the exact amounts. It is unlikely that the outcome of an investigation will be drastically different whether a dust mixture contains 80\% or 87\% dust from a specific source. Figure~\ref{PredictedProportions_OneTrace_ATKnown} shows that our model can predict, within a reasonable interval (i.e., around 5\%), the proportion of dust from an unknown source and figure~\ref{PredictedProfiles_OneTrace_ATKnown_LQ} shows that it can also extract the main characteristics of the unknown source's profile with a reasonable accuracy. 

\section{Conclusion}
Dust particles recovered from beneath the soles of an individual's shoes consist of a mixture of dust particles collected from different sources and may be indicative of the locations recently visited by that individual. In particular, this dust may reveal his presence at a location of interest, e.g., the scene of a crime. In this paper, we propose a model for the deconvolution of mixtures of dust originating from $M$ sources. Our goal is to infer the particle profiles of the $M$ sources, as well as their respective contribution to the mixture. Our overarching purpose is to enable the use and interpretation of dust evidence in order to determine, for example, if the dust recovered from under a pair of shoes contains particles originating from a given crime scene. 

We describe the profiles of each of the $M$ dust sources using a multinomial distribution over a fixed number of particle types. We use latent Dirichlet allocation (LDA) to define the probability distribution of the dust mixture. We use variational Bayesian inference (VBI) to study the source mixing proportions and particle profiles of each of the $M$ sources present in the dust mixture. Finally, we propose a method to constrain our model to learn the dust profiles of known sources using control samples collected at locations of interest (such as crime scenes, houses of suspects, etc.), while retaining the model's ability to learn the dust profiles of sources that are present in the mixtures but cannot be directly observed (such as the unknown location where a body is buried).

We test the performance of the model using real and simulated data. We find that our model is able to effectively extract the particle profiles of the sources in the mixtures present in the real data set when ``pure'' samples from all sources present in the mixtures are used to resolve them. 
Our simulations indicate that the accuracy of our model appears to be a function of the number of dust particles, the proportion of the different sources in the dust mixtures, and the magnitude of the ratios between the proportions of given particle types in the different sources. These results are very similar to the observations made regarding the well-established examination of DNA evidence in forensic science.  

We observe that our model behaves very differently depending on the constraints used for the numerical optimisation of its Dirichlet parameters. The lack of consistency of our model may be rooted in a lack of identifiability of LDA models in general. Very little has been published on the subject of identifiability of LDA models. Furthermore, models proposed in the literature are not tested using datasets with known parameters and, therefore, their accuracy cannot be assessed. This is clearly an open field for future research. 

The performance of our model in various situations needs to be extensively tested before it can be used in forensic practice. That said, it is capable of resolving mixtures of dust sources to a satisfactory level from a forensic operational perspective, and thus, of enabling forensic examiners to quantitatively support their inference of the presence of a suspect/object at a location of interest by examining dust evidence. While the transfer and examination of dust evidence was only considered as a theoretical concept by the founding fathers of forensic science, our model shows that dust particles have a great potential as a forensic tool in the near future. 

\section*{Acknowledgements}

This project was supported in part by Award No. 2016-DN-BX-0146 awarded by the National Institute of Justice (Office of Justice Programs, U.S. Department of Justice) to Stoney Forensic, Inc. (Chantilly, VA, USA), and Award No. 2014-IJ-CX-K088 awarded by the National Institute of Justice to South Dakota State University (Brookings, SD, USA). The opinions, findings, and conclusions or recommendations expressed in this paper are those of the authors and do not necessarily reflect those of the Department of Justice or the National Science Foundation. We would like to thank Dr. David Stoney for bringing up this interesting problem and providing the data.


\clearpage

\begin{appendices}

\section{Topic modelling and dust modelling: parallel terms}
\label{terms}

The following list of terms connects the vocabularies used in the topic modelling and dust modelling anecdotes of section~\ref{LDA_Dust}:

\begin{enumerate}[(a)]
	\item \textit{Dust particle}: a dust particle corresponds to a word in a topic model.
	\item \textit{Sample}: a collection of dust particles. A sample of dust corresponds to a document in a topic model. 
	\item \textit{Source}: a process or geographical area that yields dust. A source corresponds to a topic in a topic model. 
	\item \textit{Sampling Location}: a geographical area or an object where a set of samples of dust is obtained. A location may resemble an author in a topic model in the sense that both may generate samples. 
	\item \textit{Particle types}: a pre-defined set of categories that are used to classify the dust particles. A list of particle types corresponds to the vocabulary or dictionary of words in a topic model.
\end{enumerate} 

\section{Tables of notation, variables and parameters used in development}
\label{appendix:notation}

\begin{table}[H]
	\centering 
	\renewcommand{\arraystretch}{1.4}
	\caption{Table of subscripts and superscripts used to describe dust particles and dust samples \label{table:notation1}}
	\resizebox{\textwidth}{!}{
	\small
	\begin{tabular}{p{0.6cm} p{10.9cm} p{3.2cm}}
	\hline
	\em{}&\em{Description}&\em{Variable Type}\\ 
	\hline
	$l$ & Indicates which location a sample corresponds to, where $l \in \{1, \dots, L\}$ & Observed variable \\
	$s$ & Indicates which sample is being considered, where $s \in \{1, \dots, S_l\}$ & Observed variable\\ 
	$n$ & Indicates which particle is being considered, where $n \in \{1, \dots, N_{ls}\}$ & Observed variable \\ 
	$m$ & Indicates which source produced a particle , where $m \in \{1, \dots, M\}$, where $M=Q+K$ & Latent variable \\
	$t$ & Indicates which particle type is being considered, where $t \in \{1, \dots, T\}$ & Observed variable \\
	\hline
	\end{tabular}}
\end{table} 

\begin{table}[H]
	\centering 
	\renewcommand{\arraystretch}{1.5}
	\caption{Table of variables and parameters used in development \label{table:notation2}}
	\resizebox{\textwidth}{!}{
	\small
	\begin{tabular}{p{0.6cm} p{10.9cm} p{3.2cm}}
	\hline
	\em{}&\em{Description}&\em{Variable Type}\\ 
	\hline
	$\mathbf{X}$ & A set of $\sum_{l=1}^L S_l $ samples; This set includes all samples obtained from all locations (and trace objects) & Observed variable \\ 
	$\mathbf{x}_{ls}$ & The $s^{th}$ sample of dust particles from a single location $l$, such that $\mathbf{x}_{ls} := \{ x_{ls1}, x_{ls2}, \dots, x_{ls{N_{ls}}} \}$, where $N_{ls}$ is the number of dust particles in sample $s$ from location $l$ & Observed variable \\
	$x_{lsnt}$ & The particle type associated with particle $n$ from the $s^{th}$ sample from location $l$. $\mathbf{x}_{lsn}$ is an indicator vector of length $T$, such that when the $t^{th}$ position of $\mathbf{x}_{lsn}$ is equal to 1, then particle $\mathbf{x}_{lsn}$ is of type $t$ & Observed variable \\
	$\mathbf{Z}$ & An $\sum_{l=1}^L \sum_{s=1}^S N_{ls} \times M$ matrix of the sources associated with all particles from all samples across locations & Latent variable \\
	$z_{lsnm}$ & The source associated with particle $\mathbf{x}_{lsn}$; $\mathbf{z}_{lsn}$ is an indicator vector of length $M$, such that when the $m^{th}$ position of $\mathbf{z}_{lsn}$ is equal to 1, then particle $\mathbf{x}_{lsn}$ originates from source $m$ & Latent variable \\
	$\mathbf{\Theta}$ & An $\sum_{l=1}^L S_l \times M$ matrix of mixing proportions of sources for all samples across all locations & Latent variable \\
	$\boldsymbol{\mathbf{\theta}}_{ls}$ & A vector of mixing proportions associated with sample $s$ from location $l$ & Latent parameter \\
	$\theta_{lsm}$ & The proportional contribution of the $m^{th}$ source to the $s^{th}$ sample from location $l$ & Latent parameter \\
	$\mathbf{B}$ & A $M \times T$ matrix of the probabilities of observing each of the $T$ particle types at all sources $M$ for the $s^{th}$ sample & Latent parameter \\
	$\boldsymbol{\mathbf{\beta}}_{m}$ & The vector of probabilities associated with observing each of the $T$ particle types at source $m$ in sample $s$ & Latent parameter \\
	$\beta_{mt}$ & The probability of observing particle type $t$ at source $m$ in sample $s$ & Latent parameter \\
	$\mathbf{A}$ & An $L \times M$ matrix of Dirichlet distribution parameters that drive the mixing proportions of the $M$ sources at each of the $L$ locations & Latent parameter \\ 
	$\boldsymbol{\mathbf{\alpha}}_l$ & The vector of Dirichlet distribution parameters that drive the mixing proportions of the $M$ sources at location $l$  & Latent parameter \\
	$\alpha_{lm}$ & The Dirichlet distribution parameter that drives the mixing proportion of source $m$ at location $l$ & Latent parameter \\
	$\mathbf{H}$ & A $M \times T$ matrix of Dirichlet distribution parameters that drive the mixing proportions of the $T$ particle types at each of the $M$ sources & Latent parameter \\
	$\boldsymbol{\mathbf{\eta}}_m$ & The vector of Dirichlet distribution parameters that drive the mixing proportions of the $T$ particle types for source $m$ & Latent parameter \\
	$\eta_{mt}$ &The Dirichlet distribution parameter that determines the mixing proportion of particle $t$ at source $m$ & Latent parameter \\
	\hline
	\end{tabular}}
\end{table} 
\normalsize

\section{Further discussion on the M-Step}
\label{appendix:MStep}
L-BFGS-B is a limited-memory quasi-Newton method that incorporates box constraints into the optimisation process \citep{Byrd:1994, Zhu:1994}. The method serves to minimise (or, likewise, maximise) a function, $f(x_k)$, subject to the condition that $l\leq x_k \leq u$, where $l$ and $u$ represent the lower and upper bounds specified for the current iterate, $x_k$. L-BFGS-B avoids the computational cost of explicitly computing a Hessian matrix, $\nabla^2f(x_k)$, and instead, approximates $\tilde{\nabla}^2f(x_k)$ using the gradient of the function to be optimised, $\nabla f(x_k)$, and is thus efficient for large-scale problems.  The algorithm proceeds by defining a quadratic function in terms of the original function, $f(x_k)$, to be minimised, the gradient, $\nabla f(x_k)$, and a positive definite limited-memory Hessian approximation, $\tilde{\nabla}^2 f(x_k)$. Minimising this quadratic function provides an approximate solution for the next iterate, $\tilde{x}_{k+1}$, from which we can obtain a search direction. This search direction allows us to find the next iterate $x_{k+1}$. Given $x_{k+1}$, a new gradient $\nabla f(x_k)$ and limited-memory Hessian $\tilde{\nabla}^2(x_k)$ are computed, and, pending satisfaction of some convergence criterion, the next iteration begins. 

In deconvolving mixtures of dust particles, L-BFGS-B can be used to obtain updates for the matrices $\mathbf{H}$ and $\mathbf{A}$ by maximising the lower bound function $\mathscr{L}\left(q\left(\mathbf{\Theta}, \mathbf{B}, \mathbf{Z}\right)\right)$ with respect to each $\eta_{mt}$ and $\alpha_{lm}$. Clearly, the gradient plays a central role in this method, and so we use this appendix to define the gradients used in the L-BFGS-B algorithm to obtain the updates for $\mathbf{H}$ and $\mathbf{A}$. For further discussion on the L-BFGS-B method, see \citet{Byrd:1994} and \citet{Zhu:1994}.

It is convenient to begin by defining the lower bound in terms of the considered parameters. We note that the lower bound functions $\mathscr{L}_{\eta_{mt}}$ and $\mathscr{L}_{\alpha_{lm}}$ can be optimised independently, since neither function depends on  the parameters of the other:
\footnotesize
\begin{equation}
	\begin{aligned}
		\mathscr{L}_{\eta_{mk}} &= \log\Gamma\left(\sum_{j=1}^T\eta_{jt} \right)-\sum_{t=1}^T\log\Gamma\left(\eta_{mt}\right)+\sum_{t=1}^T(\eta_{mt})\left(\Psi\left(\lambda_{smt}\right)-\Psi\left(\sum_{j=1}^T\lambda_{smj}\right)\right) \\
		\mathscr{L}_{\alpha_{lm}} &= \log\Gamma\left(\sum_{j=1}^M\alpha_{lj} \right)-\sum_{m=1}^M\log\Gamma\left(\alpha_{lm}\right)+\sum_{m=1}^M\left(\alpha_{lm}\right)\left(\Psi\left(\gamma_{lsm}\right)-\Psi\left(\sum_{j=1}^M\gamma_{lsj}\right)\right).
	\end{aligned}
\end{equation}\normalsize

Indeed, specifying $\mathscr{L}_{\alpha_{lm}}$ and $\mathscr{L}_{\eta_{mt}}$ makes it straightforward to define the gradients:
\footnotesize
\begin{equation}
\begin{aligned}
\nabla \mathscr{L}_{\boldsymbol{\mathbf{\eta}}_{mt}}&=\left(\frac{\partial\mathscr{L}_{\eta_{m1}}}{\partial\eta_{m1}}, \cdots, \frac{\partial\mathscr{L}_{\eta_{mT}}}{\partial\eta_{mT}} \right) \\
\nabla\mathscr{L}_{\boldsymbol{\alpha}_l}&=\left(\frac{\partial\mathscr{L}_{\alpha_{l1}}}{\partial\alpha_{l1}}, \cdots, \frac{\partial\mathscr{L}_{\alpha_{lM}}}{\partial\alpha_{lM}}\right),
\label{gradient}
\end{aligned}
\end{equation}\normalsize

where each $\frac{\partial\mathscr{L}_{\alpha_{l\cdot}}}{\partial\alpha_{l\cdot}}$ and $\frac{\partial\mathscr{L}_{\eta_{m\cdot}}}{\partial\eta_{m\cdot}}$ is given by: 
\footnotesize
\begin{equation}
\begin{aligned}
		\frac{\partial\mathscr{L}_{\eta_{mt}}}{\partial\eta_{mt}}&= \Psi\left(\sum_{j=1}^T \eta_{mj}\right)-\Psi\left(\eta_{mt}\right)+\Psi(\lambda_{lsmt})-\Psi\left(\sum_{j=1}^T\lambda_{lsmj}\right) \\
		\frac{\partial\mathscr{L}_{\alpha_{lm}}}{\partial\alpha_{lm}}&= \Psi\left(\sum_{j=1}^M \alpha_{lj}\right)-\Psi\left(\alpha_{lm}\right)+\Psi(\gamma_{slm})-\Psi\left(\sum_{j=1}^M\gamma_{slj}\right).
\end{aligned}	
\end{equation}\normalsize

\end{appendices}

\end{document}